\documentclass[iop]{emulateapj}

\usepackage{rotating}
\usepackage{graphicx}        
\begin{document}
\title{A Biomass-Based Model to Estimate the Plausibility of Exoplanet Biosignature Gases}

\author{S. Seager\footnote{Dept. of Earth, Atmospheric and Planetary Sciences,
Massachusetts Institute of Technology, 77 Massachusetts Ave.,
Cambridge, MA, 02139.}$^{,}$\footnote{Dept. of
Physics, Massachusetts Institute of Technology, 77 Massachusetts Ave.,
Cambridge, MA, 02139}, {W. Bains$^{1,}$\footnote{Rufus Scientific}}, {R. Hu$^1$}}


\begin{abstract}
Biosignature gas detection is one of the ultimate future goals for
exoplanet atmosphere studies.  We have created a framework for linking
biosignature gas detectability to biomass estimates, including
atmospheric photochemistry and biological thermodynamics. The new
framework is intended to liberate predictive atmosphere models from
requiring fixed, Earth-like biosignature gas source fluxes. New
biosignature gases can be considered with a check that the biomass
estimate is physically plausible. We have validated the models on
terrestrial production of NO, H$_2$S, and CH$_4$, CH$_3$Cl, and
DMS. We have applied the models to propose NH$_3$ as a biosignature
gas on a ``cold Haber World'' a planet with an N$_2$-H$_2$ atmosphere,
and to demonstrate why gases such as CH$_3$Cl must have too great of a
biomass to be a plausible biosignature gas on planets with Earth or
early-Earth-like atmospheres orbiting a sun-like star. To construct
the biomass models we developed a functional classification of
biosignature gases, and found that gases (such as CH$_4$, H$_2$S,
N$_2$O) produced from life that extracts energy from chemical
potential energy gradients will always have false positives because
geochemistry has the same gases to work with as life does, and gases
(such as DMS, CH$_3$Cl) produced for secondary metabolic reasons are
far less likely to have false positives, but because of their highly
specialized origin are more likely to be produced in small quantities.
The biomass model estimates are valid to one or two orders of
magnitude; the goal is an independent approach to testing whether a
biosignature gas is plausible rather than on a precise quantification
of atmospheric biosignature gases and their corresponding biomasses.
\end{abstract}


\section{Introduction}
\label{sec-introduction}

The future detection of signs of life on exoplanets through the detection of
atmospheric biosignature gases has been a topic of long-standing
interest \citep[e.g.,][]{love1965}.  With the push to
discover exoplanets of lower and lower masses, the foundation for
biosignature gases is becoming more relevant.  The sheer variety of
exoplanets is furthermore motivating the community to take a broader
view of biosignature gases than has been conventionally considered.

\subsection{Exoplanet Biosignature Background}
\label{sec-introbiosig}

The canonical concept for the search for atmospheric biosignature
gases is to find a terrestrial exoplanet atmosphere severely out of
thermochemical redox equilibrium \citep{love1965}. Redox
chemistry\footnote{Redox chemistry adds or removes electrons from an
  atom or molecule (reduction or oxidation, respectively).} is used by
all life on Earth and is thought to enable more flexibility for
biochemistry than nonredox chemistry \citep{bain2012}. Redox chemistry
is also used to capture environmental energy for biological use. The
idea is that gas byproducts from metabolic redox reactions can
accumulate in the atmosphere and would be recognized as biosignature
gases because abiotic processes are unlikely to create a redox
disequilibrium. Indeed, Earth's atmosphere has oxygen (a highly
oxidized species) and methane (a very reduced species), a combination
several orders of magnitude out of thermodynamic equilibrium.

In practice it could be difficult to detect molecular features of two
different gases that are out of redox disequilibrium.  The Earth as an
exoplanet, for example \citep[see Fig. 8 in][]{mead2010}, has a
relatively prominent oxygen absorption feature at 0.76 $\mu$m, whereas
methane at present-day levels of 1.6~ppm has only extremely weak
spectral features. During early Earth, CH$_4$ may have been present at
much higher levels (1,000 ppm or even 1\%), as it was possibly
produced by widespread methanogenic bacteria \citep[][and references
  therein]{haqq2008}. Such high CH$_4$ concentrations would be easier
to detect, but because the Earth was not oxygenated during early
times, O$_2$ and CH$_4$ would not have been detectable concurrently
\citep[see][]{desm2002}. There may have been a short period of time in
Earth's history when CH$_4$ levels were high and before the rise of
oxygen when both could have been detected \citep{kalt2007}.

The more realistically identifiable atmospheric biosignature gas from
future remote sensing observations is a single gas completely out of
chemical equilibrium with the other known or expected atmospheric
constituents. Earth's example again is oxygen or ozone. With the oxygen
level about ten orders of magnitude higher than expected from
equilibrium chemistry \citep{kast1981, segu2007, hu2012} and has no known
abiotic production at such high levels. Although a single biosignature
gas may be all that is detectable by future exoplanet atmosphere
observations, reliance on a single biosignature gas is more prone to
false positives than the detection of two (or more) gases that are out
of equilibrium.  In the paradigm of detecting signs of life by a
single biosignature gas, we still retain the assumption that life use
chemical reactions to extract, store, and release energy, such that
biosignature gases are generated as byproducts somewhere in life's
metabolic process.

How can we decide upon the exoplanet atmosphere gases that are
identifiable as biosignature gases? Regardless of the strategy used,
only the spectroscopically active, globally-mixed gases would be
visible in an exoplanet spectrum.  Most work to date has focused on
conservative extensions of the dominant biosignature gases found on Earth,
O$_2$ (and its photochemical product O$_3$) and N$_2$O, as well as the
possibility of CH$_4$ on early Earth. Research forays into
biosignature gases that are negligible on present-day Earth but may
play a significant role on other planets has started. \citet{pilc2003}
suggested that organosulfur compounds, particularly methanethiol
(CH$_3$SH, the sulfur analog of methanol) could be produced in high
enough abundance by bacteria, possibly creating a biosignature on
other planets. CH$_3$Cl was first considered by \citet{segu2005} and
sulfur biogenic gases on anoxic planets were comprehensively
investigated by \citet{doma2011}.

A slight deviation from terracentricity is to consider Earth-like
atmospheres and Earth-like biosignature gases on planets orbiting M
stars.  \citet{segu2005} found that CH$_4$, N$_2$O, and even CH$_3$Cl
have higher concentrations and, therefore, stronger spectral features
on planets orbiting quiet M stars compared to Earth. The
\citet{segu2005} work strictly focuses on Earth's production rates for
the biosignature gases.\footnote{In effect \citet{segu2005} and others
  assume that Earth was transported as is, with its modern atmosphere,
  oceans, and biosphere, into orbit around an M-dwarf star. While this
  is a useful starting point, it is clearly a very special case.}  The
reduced UV radiation on quiet M stars enables longer biosignature gas
lifetimes and, therefore, higher concentrations to
accumulate. Specifically, lower UV flux sets up a lower atmospheric
concentration of the OH radical than in Earth's solar UV
environment. OH is the major destructive radical in Earth's atmosphere
and with less OH, most biosignature gases have longer
lifetimes. \citet{seag2012} have reviewed the range of gases and
solids produced by life on Earth.

A necessary new area of biosignature gas research will be predicting
or identifying molecules that are potential biosignature gases on
super Earth planets different from Earth. The reasons are two
fold. First, the microbial world on Earth is incredibly diverse, and
microorganisms yield a broad range of metabolic byproducts far beyond
the gases called out in exoplanet biosignature research so far. In an
environment different from Earth's, these metabolic byproducts may
accumulate to produce detectable biosignature gases different from
those on past and present Earth.  Second, while we anticipate the
discovery of transiting super Earths in the habitable zones of the
brightest low-mass stars \citep{nutz2008} and in the future Earths
from direct imaging \citep[e.g.,][]{cash2006, trau2007, laws2009}, the
prize targets around bright stars will be rare.  It follows that the
chance of finding an Earth twin might be tiny and so we must be
prepared to identify a wide range of biosignature gases.

In this paper we take a step forward to expand the possibilities for
biosignature gas detection in the future.  We provide a quantitative
framework to consider biosignature gas source fluxes of any type and
any value in any exoplanet environment, via a new biomass model
estimate that provides a physical reality plausibility check on the
amount of biomass required.  This new method liberates modelers from
assuming that exoplanet biosignature gas source fluxes are identical
to those on Earth.

\subsection{Terrestrial Biofluxes}
\label{sec-biofluxes}

We summarize terrestrial biosignature gas fluxes for later reference
as to what is a physically reasonable local ($F_{\rm field}$ in units of
mole~m$^{-2}$~s$^{-1}$) and global annual total biosignature gas flux
($\mathcal{F}_{\rm global},$ in units of Tg~yr$^{-1}$).  Biological
production of gases on Earth are limited by the availability of energy
and nutrients. We emphasize that these terrestrial biosignature gas
fluxes---which we call field fluxes ---are strictly used in this work
for consistency checks by comparison with our calculated biosignature
gas fluxes.

For Earth as a whole, the dominant energy-capture chemistry is
photosynthesis. Photosynthesis generates around $2.0 \times 10^5$~Tg
(of oxygen)~yr$^{-1}$ \citep[e.g.,][]{frie2009}. The primary carbon
production rate from photosynthesis is about $1 \times
10^5$~Tg~yr$^{-1}$ \citep{fiel1998}.

Earth has many biosignature gases beyond photosynthesis-produced
O$_2$. Some of the other biosignature gases can be produced at
relatively high flux rates, as listed in Table~\ref{tab-FieldFlux}. In
Table~\ref{tab-FieldFlux} we list the geometric mean of the maximum
field flux from one or more environmental campaigns. The main point is
that very high fluxes of biosignature gases can be generated where the
surface environment is appropriate (suitable levels of relevant
nutrients and energy sources).

We now turn to some specific examples of high terrestrial biosignature
gas fluxes. As listed in Table~\ref{tab-FieldFlux}, fluxes of some
biosignature gases (e.g., isoprene and N$_2$O) can be very large when
extrapolated from their local maximum values to a global total. In
addition to the values in Table~\ref{tab-FieldFlux}, biogenic NO$_x$
fluxes from natural (unfertilized) environments can be 10 to 30 ng (N)
m$^{-2}$~s$^{-1}$ \citep{will1992, davi1997}, which translates to a
global flux across Earth's land surface of $\sim$150 to 300
Tg~yr$^{-1}$. For environments where organic matter, water, and other
nutrients are abundant (such as swamps), flux rates of methane can
reach $10^4$ ng~(C)~m$^{-2}$~s$^{-1}$ \citep{prie1994, Dala2008},
which if scaled to a global flux would be 10 to 20 Pg~yr$^{-1}$. We
note, however, that scaling the flux from a swamp, which is rapidly
degrading biomass imported from other environments, to a global flux
is not realistic, so we do not include these methane rates in
Table~\ref{tab-FieldFlux}.

\begin{table}[h]
\begin{center}
\begin{tabular}{| l | l | l | l|}
\hline
Molecule  & Field Flux  &              Equivalent  & Global  \\
          &             &              Global Flux & Flux    \\
          & (mole m$^{-2}$~s$^{-1}$) & (Tg yr$^{-1}$) & (Tg yr$^{-1}$)   \\
\hline 
CH$_3$Cl  & 6.14$\times 10^{-12}$ & 1.5 & 2-12  \\
COS       & 1.68$\times 10^{-11}$ & 4.7 & --  \\
CS$_2$    & 2.10$\times 10^{-10}$ & 7.5 & 0.1-0.19  \\
DMS       & 3.61$\times 10^{-10}$ & 105 & 15-25  \\
H$_2$S    & 2.08$\times 10^{-10}$ & 33 & 0.2-1.6  \\
isoprene  & 8.38$\times 10^{-9}$ & 2.7$\times 10^3$ & 400-600   \\
N$_2$O    & 5.22$\times 10^{-9}$ & 1.1$\times 10^3$ & 4.6- 17  \\
NH$_3$    & --                   & --               & 10.7     \\
\hline 
\end{tabular}
\end{center}
\caption{Field fluxes from local environmental measurements for select
  biosignature gases. The geometric mean of the maximum measured field
  flux values from different studies are given.  Also listed are the
  equivalent corresponding global fluxes if the maximum field fluxes
  were present everywhere on Earth's land surface, as well as the
  actual terrestrial global flux values for comparison.  Field flux
  NH$_3$ values are not reported because on Earth free NH$_3$ is
  neglible as emission from biological systems. Global flux values for
  COS is absent because soils on average are net absorbers; Watts
  (2000) report global COS fluxes as 0.35$\pm$0.83~Tg~yr$^{-1}$.
  Field flux meaurement references are provided in the Appendix.  The
  global flux references are from \citet{sein2006}, with the exception
  of isoprene which is from \citet{arne2008}.}
\label{tab-FieldFlux}
\end{table}

Measured field fluxes, $F_{\rm field}$ as presented in the literature
vary over many orders of magnitude for the same gas species either
within a given study or among different studies. We must therefore
take an average of the literature reported $F_{\rm field}$ values and
we choose to take the geometric mean of the maximum $F_{\rm field}$
values reported in each study.  We choose to use the maximum of the
field fluxes for a given study; because the maximum represents the
ecology with the maximum number of gas-producing organisms in an
environment where they are producing gas with maximal efficiency, and
a minimum density of non-producing organisms and gas-consuming
organisms.  The huge variation in measured $F_{\rm field}$ is due to
different growth conditions, different nutrient and energy supply, and
different diffusion rates.  It is important to note that the biomass
of bioflux-producing organisms in the field is rarely measured.
Because the field fluxes are measured from an ecosystem with a range
of organisms other than the bioflux-producing organisms, in some cases
the biosignature gas of interest is consumed before it reaches the
atmosphere.  To take an average of all of the maximum field fluxes
from different studies for a given organism, we use the geometric mean
(of the maximum).

The geometric mean is the appropriate average of concentrations of
processes limited by energy. There is a log relationship between
concentration and the energy needed to drive a reaction:
\begin{equation}
\label{eq:GibbsFreeEnergy}
\Delta G = -RT \ln(K),
\end{equation}
where, $G$ is the Gibbs free energy, $R$ is the universal gas
constant, $T$ is temperature, and $K$ is the equilibrium constant. To
properly compare a set of concentrations produced by metabolism
requiring energy, the logs of the concentrations are appropriate.  For
$R_{\rm lab}$ just as for $F_{\rm field}$ we take the geometric mean
of the maxima of each study.  We choose the maximum observed rate
because it represents the closest approximation to the case where the
organisms are dependent on the gas generating reaction for the
majority of their energy. 

The bioflux produced by laboratory cultures is also relevant for
exoplanet biomass calculations, in addition to the above described
field fluxes.  We call the lab-measured metabolic byproduct production
rate per unit mass $R_{\rm lab}$, in units of
mole~g$^{-1}$~s$^{-1}$. The $R_{\rm lab}$ values of a variety of
biosignature gases are listed in Table~\ref{tab-LabRates}.  $R_{\rm
  lab}$ is used for validation of the biomass models and as input into
one of the biomass models (see \S\ref{sec-TypeIIIModel}). The $R_{\rm
  lab}$ measurements are an important complement to field measurements
as they are made on pure cultures of known mass, unlike the mixed
culture of unknown mass in the field. A summary of different flux
definitions is provided in Table~\ref{tab-FluxDefns}.

\begin{table*}[ht]
\hfill{}
\begin{center}
\begin{tabular}{| l | l | l | l| l| l|}
\hline
Molecule & Sea & Seaweed & Land Micro & Land Macro & Adopted Value \\
          & (mole g$^{-1}$~s$^{-1}$) & (mole g$^{-1}$~s$^{-1}$) & (mole g$^{-1}$~s$^{-1}$) & (mole g$^{-1}$~s$^{-1}$) & (mole g$^{-1}$~s$^{-1}$)\\
\hline
N$_2$O   & --                   & --  & 9.88$\times 10^{-10}$ & --  & 9.88$\times 10^{-10}$ \\
NO       & --                   & --  & 4.57$\times 10^{-10}$ & --  & 4.57$\times 10^{-10}$ \\
H$_2$S   & 1.00$\times 10^{-4}$ & --  & 4.69$\times 10^{-7}$  & --  & 4.51$\times 10^{-6}$ \\
CH$_4$   & 2.27$\times 10^{-8}$ & --  & 2.92$\times 10^{-6}$  & --  & 8.67$\times 10^{-7}$ \\
\hline
CH$_3$Br & 8.87$\times 10^{-12}$ & 1.04$\times 10^{-14}$  & -- & 1.23$\times 10^{-14}$ & 8.87$\times 10^{-12}$ \\
CH$_3$Cl & 6.17$\times 10^{-11}$ & 3.04$\times 10^{-15}$  & -- & 5.80$\times 10^{-12}$ & 6.17$\times 10^{-11}$ \\
COS      & 1.75$\times 10^{-16}$ & --                     & -- & 3.15$\times 10^{-14}$  & 3.15$\times 10^{-14}$  \\
CS$_2$   & 2.61$\times 10^{-14}$ & --                     & -- & --                   & 2.61$\times 10^{-14}$ \\
DMS      & 3.64$\times 10^{-7}$  & --                     & 2.45$\times 10^{-15}$  & 4.80$\times 10^{-15}$ & 3.64$\times 10^{-7}$ \\
isoprene & 4.40$\times 10^{-14}$ & 2.63$\times 10^{-16}$   & 5.61$\times 10^{-10}$ & 9.00$\times 10^{-10}$ & 9.00$\times 10^{-10}$ \\
\hline 
\end{tabular}
\end{center}
\hfill{}
\caption{Laboratory flux measurements $R_{\rm lab}$ of select
  biosignature gases in units of mole~g$^{-1}$~s$^{-1}$. The adopted
  $R_{\rm lab}$ are maximum values for Type I biosignature gases
  (first four rows) and geometric means of the maximum values for Type
  III biosignature gases (last 6 rows); see \S\ref{sec-biofluxes} for
  more details. Up to dozens of individual studies were considered.
  Blank entries have no suitable data available in the literature.
  The categories are: ``sea'' = microscopic marine species
  (phytoplankton, zooplankton and bacteria); ``seaweed'' = oceanic
  macroalgal species; ``land micro'' = microscopic land-based species;
  ``land macro'' = macroscopic land-based species. Some values were
  reported in the literature per g of dry weight. Conversion from dry
  to wet weights was performed according to the following fraction dry
  weight/wet weight: seaweed = 0.18 \citep{nico1980}; bacteria = 0.35
  \citep{brat1984, simo1989}; phytoplankton = 0.2 \citep{rick1966,
    omor1969}; fungi = 0.23 \citep{bakk1983}; lichen = 0.45
  \citep{lang1976, lars1981}; land plants (green, not woody) = 0.3
  \citep{chan1987, blac1999}.  The laboratory flux meaurement
  references are provided in the Appendix.
}
\label{tab-LabRates}
\end{table*}

The lab production rates $R_{\rm lab}$ vary by several orders of
magnitude.  The variation in lab production rates is in part due to
differences in the organisms studied in the lab, but mainly due to
different laboratory conditions, especially growth conditions
(nutrient concentration, temperature, and other environmental factors
such as whether the organisms are stressed by stirring or shaking,
non-natural light levels or spectra, or the presence or absence of
trace chemicals such as metal ions.)  For $R_{\rm lab}$ for biological
reactions based on energy extraction from the environment (defined as
Type I biosignature gases; see \S\ref{sec-TypeIdefinition}) we again
use the geometric mean as an average quantity of $R_{\rm lab}$,
because the energy released is related to the log of the concentration
of the reactants and products (see above).  For biological reactions
that do not extract energy from the environment (defined as Type III
biosignature gases; see \S\ref{sec-TypeIIIdefinition}) we use the
maximum value of $R_{\rm lab}$.  Their production rate is determined
by the ecology of the organism. Ecological factors include the
chemical environment of the species and the presence of other species,
which are rarely mimicked accurately in the laboratory. As a result
laboratory production is likely to be very substantially lower than
that in the wild. We therefore take the maximum flux found in the
laboratory measurement of Type III biosignature gas production as
being the nearest approximation to the natural flux capacity.

\begin{table}[h]
\begin{center}
\begin{tabular}{| l | l | l |}
\hline
Flux & Definition & Units \\
\hline
${\mathcal F}_{\rm global}$ &  Global & Tg yr$^{-1}$ \\
& Flux & \\
$F_{\rm source}$ & Field or & mole m$^{-2}$ s$^{-1}$ \\
& Biosignature Flux & \\
$R_{\rm lab}$ &  Lab Culture & mole g$^{-1}$ s$^{-1}$  \\
 & Flux & \\
$P_{m_e}$ & Minimal Maintenance & kJ g$^{-1}$ s$^{-1}$  \\
& Energy Rate & \\
\hline 
\end{tabular}
\end{center}
\caption{Definition of fluxes.}
\label{tab-FluxDefns}
\end{table}

\subsection{Terrestrial Biomass Densities}
\label{sec-biomassdensity}

We summarize terrestrial biomass surface densities for later reference
as to what is a reasonable biomass surface density. For our exoplanet
biomass model use and validation we require an understanding of the
range of biomass surface densities on Earth.  Based on life on Earth,
a summary overview is that a biomass surface density of 10~g~m$^{-2}$
is sensible, 100~g~m$^{-2}$ is plausible, and 5000~g~m$^{-2}$ is
possible. Real world situations are nearly always limited by energy,
bulk nutrients (carbon, nitrogen), trace nutrients (iron, etc.) or all
three.

We distinguish active biomass from inactive biomass. Active biomass is
the mass of organisms metabolizing at a sufficiently high rate to grow
(see ahead to the discussion on the microbial minimal maintenance
energy consumption rate in \S\ref{sec-Pme}). Most terrestrial
environments contain an excess of material that is biologically
derived but is not actively metabolizing. For example, the mass of
organic material in soil is 10 to 100 times greater than the mass of
actively metabolizing microorganisms
\citep[e.g.,][]{ande1989m,insa1988}. Some of this organic material is
dormant organisms but most is the remains of dead organisms (bacteria,
fungi, and plants.) In the following paragraphs we are concerned
solely with the surface density of active biomass---the biomass
actively generating byproduct gases.

We now turn to some specific examples of biomass surface densities on
Earth.
 
Photosynthesizing marine microorganisms are the dominant life over the
majority of the surface area of Earth. Their biomass is limited by
phosphate, nitrogen, iron and other micro-nutrients (because there is
no ``soil'' in the surface of the deep ocean
from which to extract micro-nutrients), and reaches 5--10
g~m$^{-2}$ on the ocean surface \citep{ishi1994, karl1991, mitc1991}. Adding
nutrients can boost the photosynthesizing marine microorganism
surface density to 50 g~m$^{-2}$ or more
\citep{bish2002,bues2004,boyd2000}.

The biomass surface densities for ocean life described above are all
in about the top 10~m of water, i.e., the well-mixed surface
zone. Nearly all of the active ocean biomass is in the top layers of the ocean,
both photosynthetic organisms and their predators. The above ocean biomass
estimates do not include the biomass in the deep ocean, where light
does not penetrate. Deep-living organisms must gain their energy
either from the small amount of biological material that falls from
the photosynthetic layer above, or from rare geochemical energy
sources such as ocean ridge or mantle hot-spot volcanic sites.  While
deep heterotrophs and hotspot geotrophs are of great conceptual
importance in our understanding of the range of environments in which
life can exist, their contribution to the total active biomass of the Earth
is not dominant. 

Microbial biofilms are limited by nutrients, energy and space. Films
on seashores, in rivers, in acid mine drainage have a huge range of
organism surface densities ranging from 0.1~g~m$^{-2}$ to
10~g~m$^{-2}$ \citep{macl1987,lawr1991,neu1997,gite2000}. Densities of
1000~g~m$^{-2}$ or more can be achieved if very high density of
nutrients are provided, as, for example, in agricultural waste water
\citep{gite2000}.

The mass of actively metabolizing microorganisms in soil is
approximately\footnote{If 100--200 g of microorganisms seems high,
  note that a 1~m$^2$ of soil 10~cm deep weighs $\sim$200~kg.}
100--200~g~m$^{-2}$ \citep{olse1987, ande1989m}. Energy-generating
nutrients are probably limiting in this case: if unlimited
energy-generating substrates are provided to fungi, as occurs, for
example, in commercial mushroom farming, densities of
$>$~20,000~g~m$^{-2}$ can be reached routinely (Shen et al. 2002).

Actively metabolizing land plant tissue has surface densities varying
from 0 to over 5000~g~m$^{-2}$, depending on the availability of
energy and nutrients.  Densities of 5000--10000 g~m$^{-2}$ of active
biomass are achieved in environments where sunlight provides unlimited
energy and nutrients are provided in excess, for example in modern
agriculture settings \citep{brer1971,hami1975}.  Densities of
100~g~m$^{-2}$ are more typical of productive grasslands.

We do not include trees or forests in our biomass density
comparisons. While forests are visually very obvious, high density
accumulations of organic carbon, nearly all of that carbon is
relatively metabolic inactive. Wood (formally, secondary xylem) acts
as a passive mechanical support for trees and a conduit for transport
of water and nutrients between the metabolically active leaves and
root surfaces. Wood produces negligible amounts of metabolic product
on its own. As most of a tree is wood, it is an inappropriate
comparison for active microbial or algal biomass.  Nevertheless, for
comparison, tree biomass densities of $\sim 6.0\times10^4$~g~m$^{-2}$,
of which ~1--5\% represents actively metabolizing green matter, are
common in mature temperate forests \citep{whit1966}.

We do not include the deep rock biosphere in our estimates of biomass
density, as (as far as is known today) crustal subsurface life has
minimal direct effect on the atmosphere.  In the last decade,
organisms have been found in deep crustal rocks that use geochemical
sources of energy for growth. The amount of this ecology is
unknown---some suggest that there is as much life in the crust as on
its surface \citep{gold1992}. However crustal life's direct impact on
the atmosphere is not obvious. Subsurface life had remained
undiscovered for so long because it does not impinge on the surface
with gases, soluble molecules, or other obvious indicators that the
subsurface organisms are present.  The subsurface organisms can only
be found by drilling into the rocks. A review of a number of studies
of microbial communities found in deep drill rock samples
\citep{pede1993}, shows that there are 10$^2$ to 10$^4$
microorganisms/gram of rock. Most studies look at bore-hole
water. Typically water from deep ($>$1~km) bore-holes contains
10$^3$--$10^5$ organisms per ml, and the water probably makes up 2-4\%
of the rock by mass (i.e., organism density in the total rock is
around 10--10$^3$ cells~g$^{-1}$). Actual biomass surface densities
will depend on how thick the inhabited rock layer is. Any
extrapolation of these figures to the possible deep rock microbial
community elsewhere on Earth, let alone on an exoplanet, must be
speculative.

Closing with a total biomass on Earth estimate, the total amount of
carbon on Earth as in cellular carbon in prokaryotes is estimated as
$3.5$--$5.5 \times 10^{14}$~kg \citep{whit1998, lipp2008}.


\subsection{A New Biosignature Approach}
\label{sec-newapproach}
The main goal of this paper is the presentation of a biomass model
estimate that ties biomass surface density to a given biosignature gas
surface source flux. The motivating rationale is that with a biomass
estimate, biosignature gas source fluxes can be free parameters in
model predictions, by giving a physical plausibility check in terms of
reasonable biomass. The new approach enables consideration of a wide
variety of both gas species and their atmospheric concentration to be
considered in biosignature model predictions. In the future when
biosignature gases are finally detected in exoplanet atmospheres, the
biomass model estimate framework can be used for interpretation.

\begin{figure*}[ht]
\begin{center}
\includegraphics[scale=.65]{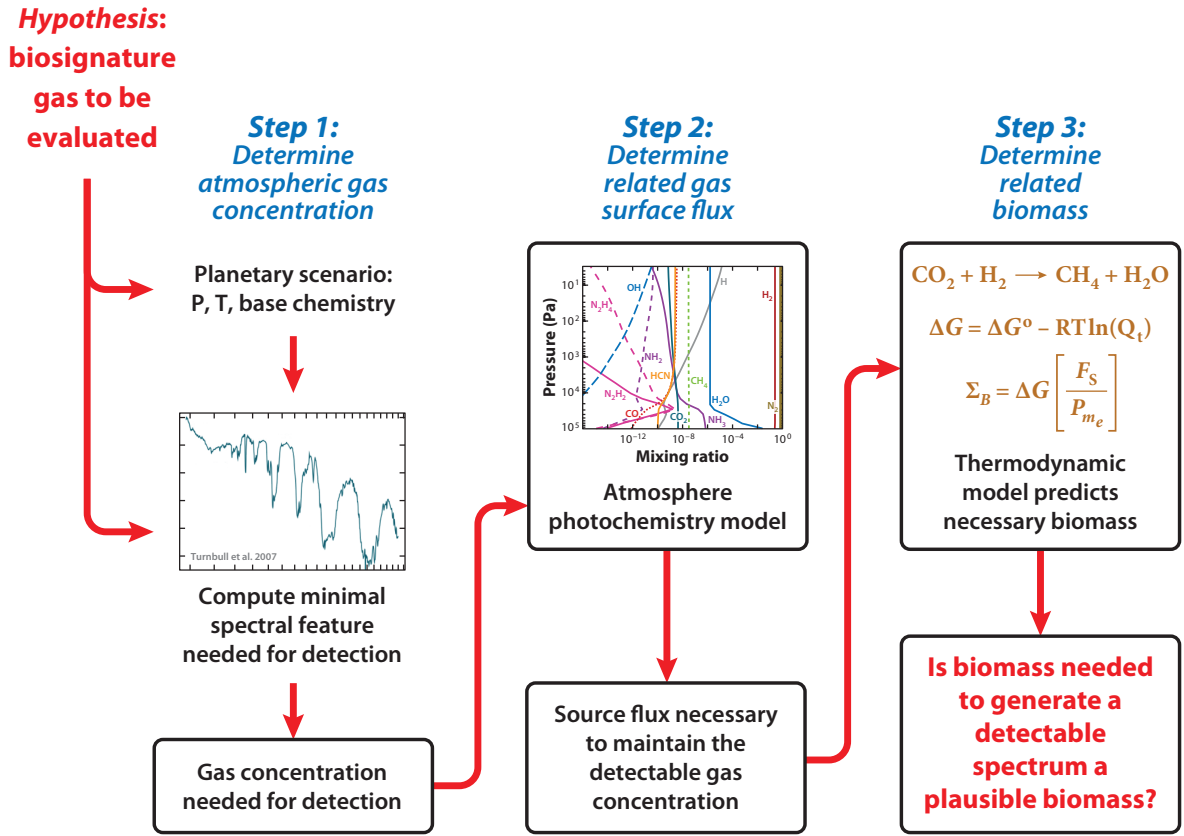}
\caption{Flow chart description of the use of biomass model estimates.
See \S\ref{sec-newapproach}.}
\label{fig:flowchart}
\end{center}
\end{figure*}

We argue that in order to explore the full range of potential
exoplanet biosignature gases, the biosignature gas source flux should
be a fundamental starting point for whether or not a biosignature gas
will accumulate in an exoplanet atmosphere to levels that will be
detectable remotely with future space telescopes. Instead, and until
now, biosignature gas fluxes are always adopted as those found on
Earth or slight deviations thereof (see \S\ref{sec-introbiosig} for
references), and could not be considered as a free parameter because
there is no first principles theoretical methodology for determining
the biosignature gas source fluxes. In lieu of a first principles
approach, we present model estimates which depend on
both the amount of biomass and the rate of biosignature gas production
per unit biomass. See Figure~\ref{fig:flowchart}.

Our proposed approach for biosignature gas studies is to:

\begin{enumerate}
\item Calculate the amount of biosignature gas required to be present 
at ``detectable''
levels in an exoplanet atmosphere from a theoretical spectrum 
(we define a detection metric in \S\ref{sec-detectionmetric});

\item Determine the gas source flux necessary to produce the
  atmospheric biosignature gas in the required atmospheric
  concentration.  The biosignature gas atmospheric concentration is a
  function not only of the gas surface source flux, but also of other
  atmospheric and surface sources and sinks
  (\S\ref{sec-photochemmodel});

\item Estimate the biomass that could produce the necessary
biosignature gas source flux (\S\ref{sec-biomassmodels});

\item Consider whether the estimated biomass surface density is physically plausible, by comparison to maximum terrestrial biomass surface density values
(\S\ref{sec-biomassdensity}) and total plausible surface biofluxes (\S\ref{sec-biofluxes}).

\end{enumerate}

We begin in \S\ref{sec-biosigclass} with a categorization of
biosignature gases into three classes, needed for the respective
biomass model estimates presented in \S\ref{sec-biomassmodels}. In
\S\ref{sec-atmandphotochem} we describe our atmosphere and
photochemistry models used to determine both the required biosignature
gas concentration for theoretical detection and the lifetimes of
biosignature gases that are produced at the planet surface. In
\S\ref{sec-results} we present our results followed by a discussion
in \S\ref{sec-discussion} and a summary in \S\ref{sec-summary}.

\section{Biosignature Gas Classification}
\label{sec-biosigclass}

A classification of biosignature gases based on their origin is needed
to develop appropriate biomass models. We make the following
definitions. Type I biosignature gases are generated as byproduct
gases from microbial energy extraction. The Type I biosignature gas
biomass model is based on thermodynamics. Type II biosignature gases
are byproduct gases produced by the metabolic reactions for biomass
building, and require energy. There is no useful biomass model for
Type II biosignature gases because once the biomass is built a Type II
biosignature gas is no longer generated.  Type III biosignature gases
are produced by life but not as byproducts of their central chemical
functions. Type III biosignature gases appear to be special to
particular species or groups of organisms, and require energy for
their production.  Because the chemical nature and amount released for
Type III biosignature gases are not linked to the local chemistry and
thermodynamics, the Type III biosignature gas biomass model is an
estimate based on field fluxes and lab culture production rates. We
further define {\it bioindicators} as the end product of chemical
reactions of a biosignature gas.

\subsection{Type I Biosignature Gas: Redox Gradient Energy Extraction Byproduct}
\label{sec-TypeIdefinition}
We define Type I biosignatures as the byproduct gases produced from
metabolic reactions that capture energy from environmental redox
chemical potential energy gradients.  Terrestrial microbes can capture
this potential energy also described as in the form of chemical
disequilibria (we also favor the term "dark energy"). Specifically,
chemotrophic organisms couple energetically favorable pairs of
oxidation and reduction half-reactions.  The disequilibria can involve
either completely inorganic compounds or can make use of organic
matter.  In fact the only clear limitations upon the types of
reactions used are that they have a negative Gibbs free energy, and
that life can make the reactions occur faster than the rate of
non-biological reactions.  In other words, Earth-based metabolic
pathways exploit chemical energy potential gradients in the form of
chemical reactions that are thermodynamically favorable but
kinetically inhibited \citep[see, e.g.,][for more details]{madi2003}.

The canonical Type I biosignature gas discussed for exoplanets is
CH$_4$ produced from methanogenesis \citep[e.g.,][and references
  therein]{desm2002}. Methanogens at the sea floor can use H$_2$
(released from rocks by hot water emitted from deep sea hydrothermal
vents (serpentinization)) to reduce CO$_2$ (available from atmospheric
CO$_2$ that has dissolved in the ocean and mixed to the bottom)
resulting in CH$_4$ and H$_2$O,
\begin{equation}
{\rm H_2 + CO_2 \rightarrow CH_4 + H_2O}.
\end{equation}
Methanogens also use molecules other than H$_2$ as reductants
(including organic molecules). For a description of volatile Type I
biosignature gases produced by Earth-based microbes (including H$_2$,
CO$_2$, N$_2$, N$_2$O, NO, NO$_2$, H$_2$S, SO$_2$, H$_2$O), see the
review by \citet{seag2012}.

On Earth many microbes extract energy from chemical energy
gradients using the abundant atmospheric O$_2$ for aerobic
oxidation,
\begin{equation} 
{\rm X + O_2 \rightarrow oxidized \hspace{0.05in} X  }.
\end{equation}
For example: H$_2$O is generated from H$_2$; CO$_2$ from organics;
SO$_2$ or SO$_4^{2-}$ from H$_2$S; rust from iron sulfide (FeS);
NO$_2^{-}$ and NO$_3^{-}$ from NH$_3$; etc.  

Turning to an exoplanet with an H-rich atmosphere, the abundant
reductant is now atmospheric H$_2$ such that
\begin{equation}
{\rm H_2 + X \rightarrow  reduced \hspace{0.05in} X }.
\end{equation}
The oxidant must come from the interior.  In other words, for chemical
potential energy gradients to exist on a planet with an H-rich
atmosphere, the planetary crust must (in part) be oxidized in order to
enable a redox couple with the reduced atmosphere. The byproduct is
always a reduced gas, because in a reducing environment H-rich
compounds are the available reductants.  Hence H$_2$S is expected from
SO$_2$; CH$_4$ from CO or CO$_2$, etc. To be more specific, oxidants
would include gases gases such as CO$_2$ and SO$_2$.
\begin{equation}
{\rm H_2 + oxidant \rightarrow CH_4 \hspace{0.05in} or \hspace{0.05in} NH_3 \hspace{0.05in} or \hspace{0.05in} H_2O}.
\end{equation}
The byproduct gases are typically those already present in
thermodynamical equilibrium. Life only has the same gases to work
with as atmospheric chemistry does.

False positives\footnote{A biosignature gas false positive is a gas
  produced by abiotic means that could be attributed to production by
  life.}  for redox byproduct gases are almost always a problem
because nature has the same source gases to work with as life
does. Furthermore, while in one environment a given redox reaction
will be kinetically inhibited and thus proceed only when activated by
life's enzymes, in another environment with the right conditions
(temperature, pressure, concentration, and acidity) the same reaction
might proceed spontaneously.  Methane, for example, is produced
geologically and emitted from mid-ocean floor ridges.  Only a reduced
gas that has accumulated to significant, unexpected levels because the
gas has a very short atmospheric life time would be a candidate
biosignature gas in an oxidized environment.  Alternatively, the
presence of reduced biosignature gases (such as CH$_4$) in an oxidized
atmosphere will stand out as candidate biosignature gases
\citep{lede1965, love1965}; but cf. comments in
\S\ref{sec-introduction} about the potential simultaneous
observability of reduced and oxidized gases.

\subsection{Type II Biosignature Gas: Biomass Building Byproduct}
\label{sec-TypeIIdefinition}
We define Type II biosignatures as byproduct gases produced by the
metabolic reactions for biomass building. On Earth these are reactions
that capture environmental carbon (and to a lesser extent other
elements) in biomass. Type II biosignature reactions are energy-consuming, and on Earth the energy comes from 
sunlight via photosynthesis. On Earth photosynthesis captures
carbon for biomass building,
\begin{equation}
{\rm H_2O + CO_2 \rightarrow CH_2O +O_2},
\end{equation}
where CH$_2$O represents sugars.  O$_2$ is Earth's most robust
biosignature gas because it is very hard to conceive of geochemical
processes that would generate a high partial pressure of oxygen in an
atmosphere with CO$_2$ in it at the Earth's atmospheric temperature,
making the probability that oxygen is a ``false positive'' signal very
low \citep{sels2002, segu2007, hu2012}. It is, however easy to explain why life
produces oxygen in an oxidized environment. In order to build biomass
in an oxidized environment, where carbon is tied up as carbonates or
CO$_2$, living organisms have to generate a highly oxidized byproduct
in order to reduce CO$_2$ to biomass.  The most plausible oxidized
species is molecular oxygen itself.  For more subtleties about why
building biomass in an oxidizing environment results in a Type II
biosignature gas that is more oxidized than the equilibrium
atmospheric components, see the detailed discussion in \citet{bain2012}.

Other potential oxidized Type II biosignature gases might include volatiles
that are oxidized forms of nitrogen (nitrogen oxides) or halogens
(molecular halogens, halogen oxides or halates; see \citet{haas2010}
regarding chloride photosynthesis), all other common elements that
could form volatile chemicals are completely oxidized in the
Earth's surface environment. The oxidized forms of nitrogen or
halogens are less likely Type II biosignature gases than oxygen
itself, as they are all more reactive (and hence damaging to
life) than molecular oxygen, require more energy to generate from
environmental chemicals, or both.

On a planet with a reduced atmosphere, we can generally state
\begin{eqnarray}
{\rm H_2O + CH_4 \rightarrow CH_2O + H_2}, \\
{\rm CO_2 + H_2O + NH_3 \rightarrow CH_2O.N + H_2},
\end{eqnarray}
etc. Here the righthand side has CH$_2$O as an approximate storage
molecule.  Because H$_2$ is the byproduct gas---already abundant in an
H-rich atmosphere--there are no useful Type II byproduct candidate
biosignature gases.  For further discussion on biosignature gases in
an H$_2$-rich atmosphere see \citep{seag2013}.

\subsection{Type III Biosignature Gas: Secondary Metabolic Byproduct}
\label{sec-TypeIIIdefinition}

We define Type III biosignatures as chemicals produced by life for
reasons other than energy capture or the construction of the basic
components of life. Type III biosignature gases have much more
chemical variety as compared to Type I or Type II biosignature gases
because they are not the products of reactions that are executed for
their thermodynamic effect out of chemicals that exist in the
environment. Rather, Type III biosignature gases have a wide variety
of functions, including defense against the environment or against
other organisms, signaling, or internal physiological control. Like
Type II biosignature gases, energy is required to generate Type III
biosignature gases.

There are a wide range of Type III biosignatures including: sulfur
compounds (e.g., DMS, OCS, CS$_2$; \citep[see][]{doma2011});
hydrocarbons; halogenated compounds (e.g., CH$_3$Cl
\citep[see][]{segu2005}, CH$_3$Br); and a variety of volatile organic
carbon chemicals (VOCs including isoprene and terpenoids).  These
products are sometimes called the products of secondary metabolism.
See \citet{seag2012} for a summary.

The most interesting aspect of secondary metabolism gas byproducts as
a biosignature class is the much more diverse range of molecules than
produced by gas products from energy extraction (the Type I
biosignature gases). Just as importantly, Type III biosignatures are
not as prone to confusion by abiotic false positives as Type I
biosignatures. As specialized chemicals, most are not naturally
occurring in the atmosphere. Because they require energy and specific
catalysis to be produced, Type III biosignature gases are unlikely to
be made geologically in substantial amounts, and so are unlikely to be
present in the absence of life.  In general, the more complicated a
molecule is (i.e., the more atoms it has) and the further from fully
oxidized or reduced the molecule is, the less are produced by
geological sources as compared to more simple molecules. For example,
volcanoes produce large quantities of CO$_2$, somewhat smaller amounts
of CH$_4$, small amounts of OCS, trace amounts of CH$_3$SH, and none
of isoprene.  The downside to Type III biosignatures is that because
they are usually such specialized compounds they typically are
produced in small quantities that do not accumulate to detectable
levels by remote sensing.

Type III biosignatures are not directly tied to the environment and
therefore could be produced by life on any exoplanet.

\subsection{Bioindicators}

Biosignature gases can be transformed into other chemical species
abiotically. The resulting product might also not be natural occurring
in a planet's atmosphere and therefore also a sign of life. We call these
abiotically altered products ``bioindicators'' and consider them a
separate subclass of each of the above three types of biosignature
gases.

O$_3$ is a canonical bioindicator derived from the Type II
biosignature O$_2$ \citep{lege1993}.  O$_3$ is a photochemical product
of O$_2$ (governed by the Chapman cycle \citep{chap1930}). As a
non-linear indicator of O$_2$, O$_3$ can be a more sensitive test of
the presence of O$_2$ under low atmospheric O$_2$ conditions
\citet{lege1993}. Other bioindicators that have been described in the
literature include ethane (a hydrocarbon compound) from biogenic
sulfur gases \citep{doma2011} and hazes generated from CH$_4$
\citep{haqq2008}.

\section{Biomass Model}
\label{sec-biomassmodels}

The main goal of this paper is a quantitative connection between
global biosignature gas source fluxes and a global biomass surface
density estimate. In this way, in models of exoplanet spectra, the
biosignature gas source fluxes can be free parameters, and checked to
be physically plausible via the biomass model estimates. Such a
plausibility check is meant to enable study of a wide variety of
candidate biosignature gases in both gas species and atmospheric
concentration to be considered.  The discussion of biosignature flux
rates and hence of biosignature detectability can thus be liberated
from the requirement of assuming Earth-like biosignature gas source
fluxes. We emphasize that we are trying to test whether a biosignature
gas can be produced by a physically plausible biomass and we are not
trying to predict what a biosphere would look like.

\subsection{Type I Biomass Model}

\subsubsection{Type I Biomass Model Derivation}

The biomass surface density for Type I biosignatures can be estimated
by conservation of energy. We may equate the required energy rate for
organism survival to the energy generation rate from an
energy-yielding reaction. The organism survival energy requirements come from an
empirical measurement of so-called minimal maintenance energy rate
that depends only on temperature \citep{tijh1993}. We describe the
minimal maintenance energy rate, $P_{m_e}$, in units of
kJ~g$^{-1}$~s$^{-1}$ (i.e., power per unit mass)\footnote{We use g as
  a proxy for g of wet weight.} later in this section. The energy yield
rate comes from the Gibbs free energy of the energy-yielding reaction times the
rate at which a group of organisms processes the reaction. The Gibbs
free energy of the reaction is denoted by $\Delta G$, in units of
kJ~mole$^{-1}$. The metabolic byproduct gas production rate per unit mass
is described by $R$ in units of mole~g$^{-1}$~s$^{-1}$.
The conservation of energy per unit mass and time is then described by\begin{equation}
\label{eq:bio1}
P_{m_e} = \Delta G \hspace{0.005in} R.
\end{equation}
The equation tells us that under the assumption that the energy yield
$\Delta G$ is used only for maintenance, the byproduct gas
production rate per unit mass $R$ can be constrained if $P_{m_e}$ is
known. The byproduct gas production rate is what we have been calling
the biosignature gas surface flux.

The biomass surface density is the parameter of interest and so we 
breakdown $R$ into a biomass surface density $\Sigma_B$
and a biosignature gas source flux $F_{\rm source}$. 
\begin{equation}
\label{eq:bio2}
R = \frac{F_{\rm source}}{\Sigma_B},
\end{equation}
where $\Sigma_B$ is the biomass surface density in g~m$^{-2}$. The
biosignature gas source flux $F_{\rm source}$ (in units of mole
m$^{-2}$ s$^{-1}$) describes the surface flux emitted as the metabolic
byproduct and is also used as an input in a computer model of an
exoplanet atmosphere. An important point for exoplanet atmospheres is
that $F_{\rm source}$ can not be directly converted into a detectable
gas concentration that makes up a spectral feature---any source flux
coming out of a planet surface is usually modified by atmospheric
chemical reactions including photochemical processes. In atmosphere
modeling, a photochemistry model is needed to translate the source
flux into the amount of gas that accumulates in an exoplanet
atmosphere. (False positives in the form of geologically-produced
source fluxes must be also be considered; see
\S\ref{sec-falsepositives}.)

The biomass estimate follows from equations~(\ref{eq:bio1}) and
(\ref{eq:bio2}),
\begin{equation}
 \label{eq:bio3}
\label{eq:TypeIBiomassModel}
\Sigma_B \simeq \Delta G \left[ \frac{F_{\rm source}}{ P_{m_e}} \right].
\end{equation}
The free parameter in this biomass estimate equation is the
biosignature gas source flux $F_{\rm source}$, because the Gibbs free
energy is known and minimal maintenance energy rate is empirically
adopted (\S\ref{sec-Pme}). A caveat is that both $\Delta G$ and
$P_{m_e}$ depend on temperature.

$\Sigma_B$ is an apparent minimum biomass surface density estimate
because $F_{\rm source}$ may be weakened to a net biosignature gas
surface flux if some of the gas is consumed by other organisms. See
\S\ref{sec-ecologycontext} for a discussion.

We review the point that $\Delta G$ depends on gas concentration. The
energy available to do work depends on the concentration of both the
reactants and products via
\begin{equation}
\label{eq:realdeltag}
\Delta G= \Delta G_0 +RT \ln (Q_t).
\end{equation}
Here $\Delta G_0$ is the ``standard free energy'' of the system
(equation~(\ref{eq:GibbsFreeEnergy})), i.e., the
free energy available when all the reactants are in their standard
state, 1 molar concentration (for solutes) or 1 atmosphere pressure
(for gases). $R$ in this context is the universal gas constant and
$T$ is temperature. The reaction quotient $Q_t$ is defined as
\begin{equation}
Q_t = \frac{[A]^n [B]^m}{[X]^o[Y]^p},
\end{equation}
for the reaction
\begin{equation}
oX + pY \rightarrow nA + mB,
\end{equation}
where, e.g., $[X]$ is the concentration or partial pressure of species
$X$.  In general, care must be taken for the Type I biomass
calculations described in this paper as relates to the appropriate
$\Delta$G. Most of our $\Delta G$ are taken from \citet{amen2001}.

The biosignature gas source flux, $F_{\rm source}$ can now be used as
a free parameter in exoplanet model atmosphere calculations, whereby,
again, the purpose of the biomass estimate from equation~(\ref{eq:bio3}) is
to ensure the biomass surface density underlying the source flux is
physically reasonable.

\subsubsection{The Minimal Maintenance Energy Consumption Rate, $P_{m_e}$}
\label{sec-Pme}
We now turn to discuss the microbial minimal maintenance energy consumption
rate, $P_{m_e}$, in more detail. Although $P_{m_e}$ is not yet a
familiar quantity in exoplanets research, it is both measured
empirically and tied to thermodynamics. $P_{m_e}$ is a
fundamental energy flux central to the biomass estimate.  $P_{m_e}$ is
the minimal amount of energy an organism needs per unit time to
survive in an active state (i.e., a state in which the organism is
ready to grow).  An empirical relation has been identified by
\citet{tijh1993} that follows an Arrhenius law
\begin{equation}
\label{eq:Pme}
P_{m_e} = A \exp\left[ \frac{-E_A}{RT} \right].
\end{equation}
Here $E_A = 6.94 \times 10^4$~J~mol$^{-1}$ is the activation energy,
$R$ = 8.314~J~mol$^{-1}$ K$^{-1}$ is the universal gas constant, and
$T$ in units of K is the temperature. The constant $A$ is $3.9 \times
10^7$~kJ~g$^{-1}$~s$^{-1}$ for aerobic growth and $2.2 \times
10^7$~kJ~g$^{-1}$~s$^{-1}$ for anaerobic growth \citep{tijh1993}.
Here per g refers to per g of wet weight of the organism.  Note that
we have explicitly converted from Tijuis' $P_{m_e}$ units of
kJ~(mole-C)$^{-1}$~yr$^{-1}$ in bacterial cells to kJ~g$^{-1}$~s$^{-1}$
per organism by dividing the $P_{m_e}$ values by a factor of 60
(molecular weight of carbon (= 12) $\times$ the ratio of dry weight to carbon
(=2) $\times$ the ratio of wet weight to dry weight in bacteria (=2.5)).

The $P_{m_e}$ maintenance energy rate equation
(equation~(\ref{eq:Pme})) is species independent \citep{tijh1993} and
also applicable for different microbial culture systems
\citep{hard1997}. The equation is not intended to be very precise, the
confidence intervals are 41\% and 32\% for aerobic and anaerobic
growth respectively \citep{tijh1993}.

$P_{m_e}$ as measured is not strictly a minimal energy
requirement. $P_{m_e}$ is in fact the minimal energy needed for a
bacterial cell to keep going under conditions under which it is
capable of growth. The $P_{m_e}$ is measured during growth, and is
extrapolated to growth = 0. This extrapolated $P_{m_e}$ is
nevertheless not the same as ``maintenance energy'' for non-growing
cells. Growing cells have a variety of energy-required mechanisms
``turned on'' which non-growing cells will turn off to save energy,
such as the machinery to make proteins, break down cell walls and so
on. $P_{m_e}$ as a maintenance energy rate therefore separates out the
baseline energy components from the energy needed to actually build
biomass.  $P_{m_e}$ is therefore the minimal energy needed to maintain
active biomass\footnote{We note that this active biomass may be
  accompanied by a much larger mass of dormant organisms (and an even
  larger mass of dead ones, as in terrestrial soils).  However dormant
  and dead organsims will not be significant generators of
  biosignature gases, and so we are not interested in them for our
  present study. Equally, we are not interested in dormant organisms'
  lower energy requirements.}. See \citet{hoeh2004} for a more
detailed review of the different types of ``maintenance energy'' and
their relationship to organism growth.

The $P_{m_e}$ (equation~(\ref{eq:Pme})) is an Arrhenius equation and
it is natural to ask why the microbial maintenance energy rate follows
Arrhenius' law.  Organisms use energy for repair and replacement of
damaged molecular components. Molecular damage is caused by
non-specific chemical attack on the components of the cell by water,
oxygen, and other reactive small molecules. The rate of such reactions
is no different than other chemical reactions---well described by an
Arrhenius equation. In aggregate, therefore, the overall rate of
breakdown of the macromolecular components of the cell is expected to
follow Arrhenius' law. Arrhenius' law describes chemical reaction
rates (indeed any thermally activated process) and has two
reaction-specific parameters $A$ and $E_A$. For two stable molecules
to react, chemical bonds need to be broken. $E_A$, the activation
energy of a reaction, represents the energy needed to break the
chemical bonds. In uncatalyzed reactions, $E_A$ comes from the thermal
energy of the two reacting molecules, which itself follows from the
Boltzmann velocity distribution. The probability that any two
colliding molecules will have a combined energy greater than $E_A$, is
$\exp[-E_A/RT]$. The parameter $A$ is an efficiency factor that takes
into account that molecules have to be correctly oriented in order to
react.

\subsubsection{Type I Biomass Model Validation}
\label{sec-TypeIBiomassValidation}
Tests of the biomass model for Type I biosignature gases aim to
both validate the model and understand the intended range of model
accuracy. Because our end goal is to estimate 
whether the flux of gas necessary to generate a spectral signature
is plausible, we aim only for an order of magnitude estimate of the 
biomass that is producing the biosignature gas of interest.

The first test is to check our basic assumption of conservation of
energy in equation~(\ref{eq:bio1}): that the maintenance energy rate
($P_{m_e}$) is approximately equal to the redox energy yield rate
($R_{\rm lab} \Delta G$), via lab measured rate values. We consider
the biosignature gases and corresponding reactions described below and
compare the maintenance energy rate to the redox energy yield rate,
along with the values for $R_{\rm lab}$, $\Delta G$, and $P_{m_e}$ at
the temperature, $T$, considered (each of these three quantities are
temperature-sensitive).  To validate using equation~(\ref{eq:bio1}) we
have averaged the validation results from different literature
studies. In other words, we used the appropriate concentration, pH,
and temperature for the $\Delta G$ and the appropriate temperature for
$P_{m_e}$ with the validation result shown in the last column in
Table~\ref{tab-TypeIValidation}.  To provide overview values for each
individual parameter, Table~\ref{tab-TypeIValidation} also shows
averaged values for each of $R_{\rm lab}$ and $\Delta G$.

We consider four different Type I biosignature gas-generating
reactions. These reactions are selected because they involve the
reaction of geochemically available starting materials, have well
characterized microbial chemistry, and for which sufficient $F_{\rm
  field}$ and $R_{\rm lab}$ measurements are available.

The first reaction is ammonia oxidation to nitrogen oxides,
described by
\begin{eqnarray}
{\rm 2 NH_4^+ + 2 O_2 \rightarrow N_2O + 3 H_2O + 2H^+} \\ \nonumber
{\rm 4 NH_4^+ + 5 O_2 \rightarrow 4 NO + 6 H_2O + 4H^+}.
\end{eqnarray}
We note that
the oxidation of ammonia is only a relevant route for production of
N$_2$O in an environment containing molecular oxygen. In both
laboratory systems and real ecosystems, organisms oxidize ammonia to
N$_2$O and to NO at the same time, the ratio depending on oxygen
availability and other environmental factors. To validate our
estimates of gas flux based on energy requirements, we therefore have
to account for an organism's production of N$_2$O {\it and} NO, as the
production of both of these gases contributes materially to the
organism's energy budget. For ammonia oxidation
we summed the $R_{\rm lab} \Delta G$ for NO and N$_2$O generation
for each experiment and
calculated the geometric mean of those summed values.

Our second, third, and fourth validation examples are for H$_2$S, a
gas produced by many biological reactions.  As examples, we choose the
reduction of elemental sulfur (at two different temperatures),
\begin{equation}
{\rm H_2 + S \rightarrow H_2S},
\end{equation}
and the disproportionation of thiosulfate,
\begin{equation}
{\rm S_2O_3^{2-} + H_2O \rightarrow HSO_4^{2-} + H_2S}.
\end{equation}
These two reactions can use geochemically produced substrates, and
hence are not dependent on pre-existing biomass.
 
We choose as a fifth example methanogenesis, via the reduction of
CO$_2$ by H$_2$ to produce CH$_4$. Methanogenesis is a key
energy-capturing reaction in hydrothermal environments, and a reaction
which relies only on geochemical inputs,
\begin{equation}
{\rm CO_2 + 4H_2 \rightarrow 2H_2O + CH_4}.
\end{equation}
Methanogenesis is discussed at length in sections
\S\ref{sec-methanogenesis} and \S\ref{sec-Mars}.

\begin{table*}[ht]
\begin{center}
\begin{tabular}{| c | c | c | c| c| c| c|}
\hline
Gas &$R_{\rm lab}$ & $\Delta G$ & $T$ & $P_{m_e}$ & Approx. $\frac{R_{\rm lab} \Delta G}{P_{m_e}}$ & $\frac{R_{\rm lab} \Delta G}{P_{m_e}}$ \\
    & [mole/(g s)] & $\left[\frac{\rm kJ}{\rm  mole}\right]$ & [K] & [kJ/(g s)]& & \\         
\hline
N$_2$O   & 3.36$\times 10^{-8}$   & 443.7 & 302 & --  & -- & --\\
NO       & 5.36$\times 10^{-9}$   & 355.9 & 302 & --  & -- & --\\
N$_2$O/NO & 8.73$\times 10^{-8}$  & -- & 302 & 2.2$\times 10^{-5}$  & 3.2 & 0.62\\
\hline
H$_2$S a & 8.17$\times 10^{-5}$   & 27.2  & 338 & 4.2$\times 10^{-4}$ & 5.3 & 5.2\\
\hline
H$_2$S b & 6.57$\times 10^{-3}$    & 13.4  & 374 & 4.8$\times 10^{-3}$ & 19 & 19\\
\hline
H$_2$S c & 1.13$\times 10^{-6}$   & 31.8  & 308 & 3.8$\times 10^{-5}$ & 0.95 & 4.8 \\
\hline
CH$_4$   & 4.7$\times 10^{-5}$    & 191.8 & 338 & 4.2$\times 10^{-4}$ & 21 & 21\\
\hline 
\end{tabular}
\end{center}
\caption{Type I biomass model validation for select biosignature
  gases. The maintenance energy production rate $P_{m_e}$ should be
  comparable to the lab production rate fluxes times the free energy
  $R_{\rm lab} \Delta G$. Averaged values for each literature study
  are given for $R_{\rm lab}$, $\Delta_G$, and $P_{m_e}$ ($T$ is
  within a few degrees for each row), with the approximate validation
  using these averages givein in column 6. The actual validation
  given in column 7 is an average of individual values (not shown) of
  $R_{\rm lab}$, $\Delta_G$, and $P_{m_e}$.
The biosignature gas producing reactions are listed in the text but
can be summarized as: N$_2$O: produced via ammonia oxidation; H$_2$S a
and b sulfur reduction with H$_2$; H$_2$S c: S$_2$O$_3$
disproportionation to H$_2$S; CH$_4$ produced via methanogenesis (see
text in \S\ref{sec-TypeIBiomassValidation}) for details.}  \normalsize
\label{tab-TypeIValidation}
\end{table*}

The validation results---that the lab-based redox energy yield rate
compares to the maintenance energy rate within an order of
magnitude---are shown in Table~\ref{tab-TypeIValidation} (right-most
column). The results show a reasonable confirmation of our application
of the minimal maintenance energy concept to biosignature gas
production rates, to within about an order of magnitude with one
exception.  An order of magnitude is expected because of uncertainties
in the individual factors: $P_{m_e}$ a factor of 2; the gas flux
typically a factor of 2; the biomass measurement and conversion values
a factor of 2).

We pause to discuss the relevance of validating the $P_{m_e}$ equation
against lab production rates, given that \citet{tijh1993} already used
laboratory measurements in the original paper. The \citet{tijh1993}
equation for $P_{m_e}$ (equation~(\ref{eq:Pme})) was developed from
studies in which all the inputs and outputs from energy metabolism
were completely characterized, so that the energy balance of the
organisms could be calculated exactly. The organisms' growth rate was
also controlled, so that the energy consumption at zero growth
($P_{m_e}$) could be inferred directly from the data. In our
application we wish to infer the biomass from a single measure of gas
output, from organisms whose rate of growth is not known. We therefore
needed to validate that such an extrapolation of the application of
the $P_{m_e}$ concept is valid.

Based on the lab and $P_{m_e}$ comparison in
Table~\ref{tab-TypeIValidation}, the rate of production of gas in
growing cultures of organism in the laboratory is higher than that
predicted by the $P_{m_e}$ calculations. This is expected. $P_{m_e}$
is the minimal maintenance energy---the energy needed to maintain the cell in
a state ready to grow. Actual growth requires additional energy to
assemble cell components. This extra energy demand in turn requires
that the cell produce more Type I metabolic waste products per unit
mass than is expected from the $P_{m_e}$ calculations. The amount of
the excess will depend on specifics of the growth conditions (e.g.,
what nutrients are supplied to the organisms), the organisms growth
rate, and specifics of its metabolism. Thus
Table~\ref{tab-TypeIValidation} is consistent with our model, showing
that organisms use at least the $P_{m_e}$ of energy in cultures
capable of active growth.
 
Now we comment more specifically on the actual validation numbers in
Table~\ref{tab-TypeIValidation}. Two of the test validation results
are too high, at a factor of 20 when they should be close to
unity. Based on this high value and reasons described further below,
the Type I biomass model should only be used for temperatures below
$\sim$343~K, because the \citet{tijh1993} $P_{m_e}$ equation was
derived from measurements taken between 283 and 338~K (with one
measurement at 343~K). Both of the anomalously high values in
Table~\ref{tab-TypeIValidation} are for cultures grown at the upper
end of this temperature range. It is possible that at such extreme
temperatures organisms require more energy for stress and damage
response than predicted from culture at lower temperature.  We note
that the deviation of $P_{m_e}$ as compared to $\Delta G R_{\rm lab}$
may also be reflected in the temperature dependencies: $P_{m_e}$
follows an exponential with $T$ whereas $\Delta G$ changes linearly
with $T$ (equation~(\ref{eq:GibbsFreeEnergy})).

As a second test of the Type I biomass model we check the biomass
estimate equation~(\ref{eq:TypeIBiomassModel}), specifically that the
quantity of interest, the biomass surface density ($\Sigma_B$) is
reasonable based on the field fluxes and $\Delta G$ and $P_{m_e}$. In
other words, for this second test we ask if the surface biomass
(estimated via equation~(\ref{eq:TypeIBiomassModel})) is reasonable by
comparison with Earth-based biomass surface densities for the
microbial redox energy equation and environment in question. For the
N$_2$O and H$_2$S examples given above, we find biomass surface
densities are below 0.0024 g~m$^{-2}$, as shown in
Table~\ref{tab-TypeIValidationb}, well within a plausible biomass
surface density (\S\ref{sec-biomassdensity}).

\begin{table}[ht]
\begin{center}
\begin{tabular}{| l | l | l | l| l| l|}
\hline
Gas & $P_{m_e}$ & $\Delta G$ & $T$ & $F_{\rm field}$ & $\Sigma_B$ \\
    &[kJ/(g s)] & [kJ/mole] & [K] & [mole/(m$^2$s)] & [g/m$^2$]  \\         
\hline
N$_2$O/NO & 4.1$\times 10^{-5}$   & -472.0 & 303 & 5.22$\times 10^{-9}$  & 2.4$\times 10^{-2}$  \\
\hline
H$_2$S a & 1.1$\times 10^{-3}$   & -46.7  & 338 &  2.08$\times 10^{-10}$  & 9.3$\times 10^{-6}$  \\
\hline
H$_2$S b & 1.1$\times 10^{-2}$    & -48.3  & 373 & 2.08$\times 10^{-10}$  & 9.4$\times 10^{-7}$  \\
\hline
H$_2$S c & 9.5$\times 10^{-5}$   & -23.9  & 307 & 2.08$\times 10^{-10}$  & 5.1$\times 10^{-5}$  \\
\hline
\end{tabular}
\end{center}
\caption{Type I biomass model validation for biomass surface density
  for select biosignature gases. The biomass surface density is
  computed by the Type I biomass model equation~(\ref{eq:bio3}) using
  the geometric means of the maximum values of the field fluxes
  $F_{\rm field}$. The biomass surface density should be reasonable as
  compared to terrestrial values described in
  \S\ref{sec-biomassdensity}}.
\label{tab-TypeIValidationb}
\end{table}

We did not validate CH$_4$ for surface biomass density because maximum
local field fluxes of methane are not meaningful for comparison with
other gas fluxes in our analysis. Extremely high CH$_4$ fluxes can be
generated from anaerobic biomass breakdown (fermentation), but these
represent the rapid breakdown on biomass that has been accumulated
over much wider areas and over substantial time. As an extreme
example, sewage processing plants can generate substantial methane,
but only because they collect their biomass from an entire city. This
flux does not therefore represent a process that could be scaled up to
cover a planet.

We also did not validate the biosignature gas NH$_3$ because no
natural terrestrial environment emits detectable amounts of ammonia on
a global scale. NH$_3$ is sometimes generated by the breakdown of
biomass, especially protein-rich biomass or nitrogen-rich excretion
products.  But NH$_3$ represents a valuable source of nitrogen, which
is taken up rapidly by life. Because ammonia is very soluble in water,
any residual NH$_3$ not taken up by life remains dissolved in water
and does not generate any significant amount of NH$_3$ gas in the
atmosphere.
 
We conclude this subsection by summarizing that the Type I biomass
model is a useful estimate to about an order of magnitude. This is
validated from our use of the minimimum maintenance energy $P_{m_e}$
as compared to lab flux values ($\Delta G R_{\rm lab}$). We also
showed that a reasonable biomass surface density is derived using the
field flux values in the main Type I biomass equation
(equation~(\ref{eq:bio3})).

\subsection{Lack of Type II Biosignature Biomass Model}

We do not propose a biomass model for Type II biosignature gases and
here we explain why.  Type II biosignature gases are
produced as a result of biomass building. Once the biomass is built,
there is no further Type II biosignature gas produced, to a reasonable
approximation.

If one wanted to estimate a Type II biosignature gas flux, one would
have to estimate the turnover rate of the biomass, which itself
depends on seasonality, burial rates, predation, fire clearance, and
many other factors. If a turnover rate, $Tr$ could be determined, then
the flux of a Type II biosignature gas would be $Tr \times s$, where
$s$ is the stoichiometrically determined amount of biosignature gas
needed to generate a gram of biomass. For plants, for example
($\sim$80\% water, $\sim$20\% dry weight; of that dry weight 45\% is
carbon), the stoichiometry of carbon fixation is 
\begin{equation} 
{\rm CO_2 + H_2O \rightarrow CH_2O + O_2},
\end{equation}
(in other words one mole of carbon fixed gives one mole of oxygen released),
 and so $s = 5.6 \times 10^{-3}$~moles~O$_2$~g$^{-1}$ wet weight of plant.

For our present purposes, in the absence of any good framework for
estimating exoplanet $Tr$, we omit biomass models for Type II
biosignature gases.

\subsection{Type III Biosignature Biomass Estimates}
\label{sec-TypeIIIModel}

Type III biosignature gases have no physically-based biomass model
because Type III biosignatures are not linked to the growth or
maintenance of the producing organism.  Because the amount of
biosignature gas produced is arbitrary from the point of view of its
overall metabolism, there can be no quantitative biomass model for
Type III biosignature gases.  

We therefore instead construct a biomass estimate from a framework
based on terrestrial Type III biosignature gas fluxes. While tying
the biomass estimate to the specifics of terrestrial metabolism is
unsatisfactory, it is still more general than the conventional
adoption of Earth-like environmental flux rates which assume both
terrestrial metabolism and terrestrial ecology. With our Type III
biomass estimate approach, we can scale the biosignature gas source
flux to different biomass densities that are not achieved on Earth.

\subsubsection{Type III Biomass Estimate}
In lieu of a quantitative, physical biomass model, we adopt a
comparative approach for a biomass estimate. We use the biosignature
source fluxes and production rates of Type III metabolites of
Earth-based organisms.

We estimate the biomass surface density by taking the biosignature gas
source flux $F_{\rm source}$ (in units of mole m$^{-2}$ s$^{-1}$)
divided by the mean gas production rate in the lab $R_{\rm lab}$ (in
units of mole~g$^{-1}$~s$^{-1}$), as in equation~(\ref{eq:bio2}),
\begin{equation}
\label{eq:TypeIIIBiomassModel}
\Sigma_B \simeq \frac{F_{\rm source}}{R_{\rm lab}}.
\end{equation}
Recall that the source flux is measured in the field on Earth (and in
that context called $F_{\rm field}$) but assumed or calculated for
exoplanet biosignature gas detectability models (and called $F_{\rm
  source}$).  See Table~\ref{tab-FieldFlux} for a list of select Type
III field fluxes and Table~\ref{tab-LabRates} for a list of select
Type III lab rates.  As described in \S\ref{sec-biofluxes} we take the
geometric mean of the maximum for the Type III $R_{\rm lab}$ rates 
$F_{\rm field}$ values from different studies.

The caveat of the Type III biomass estimate explicitly assumes that
the range of $R$ for life on exoplanets is similar to that for life in
Earth's lab environment.

\subsubsection{Type III Biomass Estimate Validation}

We now turn to a validity check of the Type III biomass estimate.  To
validate we compare the flux rates of Type III biosignature gases
observed in the field ($F_{\rm field}$) with the production rate of
Type III gases from laboratory culture ($R_{\rm lab}$) of pure
organisms.  The field rates give a flux per unit area, and the
laboratory rates give a flux rate per unit mass.  Also,
the lab rates are from single species whereas
the field rates are from an ecology. We wish to confirm
that comparison of the two predicts a physically plausible biomass per
unit area to explain the field flux. We use the values for $F_{\rm
  field}$ and $R_{\rm lab}$ as given in
Table~\ref{tab-TypeIIIValidation}.

\begin{table}[h]
\begin{center}
\begin{tabular}{| l | l | l | l|}
\hline 
Molecule & $F_{\rm field}$ & $R_{\rm lab}$& $\Sigma_B = \frac{F_{\rm field}}{R_{\rm lab}}$ \\ 
\hline
CH$_3$Cl   & 2.90$\times 10^{-12}$ & 6.17$\times 10^{-11}$ & 0.0047 \\ 
COS     & 1.68$\times 10^{-11}$ & 2.70$\times 10^{-14}$ & 620 \\ 
CS$_2$     & 3.96$\times 10^{-12}$ & 2.61$\times 10^{-14}$ & 150 \\ 
DMS     & 5.83$\times 10^{-12}$ & 3.64$\times 10^{-07}$ & $1.6\times10^{-5}$ \\
isoprene & 8.38$\times 10^{-9}$  & 9.00$\times 10^{-10}$ & 9.3 \\
\hline 
\end{tabular}
\end{center}
\caption{Type III biomass estimate validation. The biomass surface
  density, $\Sigma_B$ as generated from the Type III biomass in
  equation~(\ref{eq:TypeIIIBiomassModel}). Here $F_{\rm field}$ are
  Earth values---the geometric mean of the maximum fluxes from values
  reported in the literature, taken from
  Table~\ref{tab-FieldFlux}. $R_{\rm lab}$ are Earth lab values, the maximum fluxes from literature studies, and are taken from
  Table~\ref{tab-LabRates}.}
\label{tab-TypeIIIValidation}
\end{table}

Most of the Type III biomass surface density values (as shown in
Table~\ref{tab-TypeIIIValidation}) are well within the values of
biomass density seen in natural ecosystems
(\S\ref{sec-biomassdensity}). We can therefore say that using
laboratory fluxes is a reasonable way to approximately estimate the
biomass necessary to generate biosignature gas fluxes. We emphasize
approximate, because the biomass densities are somewhat high for Type
III organism biomass densities.

The biomass surface density validation for COS is somewhat large at
622 g~m$^{-2}$.  The problem with COS is that it is both given off and
absorbed by ecosystems, often by the same ecosystem at different
times. The net field flux may therefore be poorly defined, perhaps
representing the release of stored gas. In addition, COS is usually
only produced by organisms in response to attack by other organisms
such that COS is produced in large quantities in soils but produced in
much smaller quanaties in lab cultures. The same argument applies to
CS$_2$ and also gives rise to a somewhat large biomass. In any case,
COS and CS$_2$ are examples of how the Type III biomass model based on
scaling is approximate only.

The biomass surface density validation for DMS is much lower than the
other Type III biomass estimates in
Table~\ref{tab-TypeIIIValidation}. All of the molecules in the table
except DMS are produced at a cost to the organism for carbon and
energy in order to perform specific signaling or defense
functions. DMS, in contrast, is a product of the consumption of DMSP.
DMSP is produced in response to stress and then is broken down
enzymatically to DMS by zooplankton. In the lab DMSP is often fed to
the phytoplankton at a level unavailable in the natural environment,
and the phytoplankton consume the DMSP at a very high rate likely
leading to the high lab DMS production rates, and hence the low
biomass surface density estimate.

\section{Atmosphere and Photochemistry Model}
\label{sec-atmandphotochem}

A model for atmospheric chemistry is required to connect the
concentration of a biosignature gas in the atmosphere as required for
detection to the biosignature source flux at the planetary surface.
(In turn we use the biosignature source flux to estimate the biomass
through the models introduced in \S\ref{sec-biomassmodels}.)  The
focus on chemistry is critical, because atmospheric sinks that destroy
the putative biosignature gas are critical for the gas lifetime and
hence accumulation in the planetary atmosphere.

\subsection{Photochemistry Model}
\label{sec-photochemmodel}
We aim to calculate the source flux $F_{\rm source}$ or in
photochemistry model jargon, the production rate $P$ of a gas species
of interest. The production rate is tied to the loss rate $L$ and the
steady state gas concentration $[A]$, via
\begin{equation}
\label{eq:photochem1}
P = L[A].
\end{equation}
The source flux is described by
\begin{equation}
F_{\rm source} = \int_z P(z) + \Phi_{\rm dep},
\end{equation}
where $z$ is altitude and $\Phi_{\rm dep}$ is the deposition flux
described later below.

The derivation of the production rate (source flux) equation
is as follows. In steady state,
\begin{equation}
\label{eq:photobasic}
\frac{d[A]}{dt} = P - L[A] = 0,
\end{equation}
where ${\rm[A]}$ is the number density of species $A$ (in units
of molecule~m$^{-3}$), $P$ is the production rate of species $A$ (in
units of molecule m$^{-3}$~s$^{-1}$), and $L$ is the loss rate of
species $A$ in (in units of s$^{-1}$).
By rearranging equation~(\ref{eq:photobasic}) we have $ {\rm
  P} = L [A]$. We also note that the loss rate is often described
by its inverse, the lifetime of an atmospheric gas,
\begin{equation}
\label{eq:tforlossrate}
t = \frac{1}{L}.
\end{equation}

The loss rate can be described in more detail. The loss rate can be
due to reactions with other species $B$, as in
\begin{equation}
\label{eq:lossrate1}
L[A] = K_{AB}[A][B],
\end{equation}
where $K_{AB}$ is the second order reaction rate in units of m$^3$ molecule$^{-1}$
s$^{-1}$. Values of $K_{AB}$ are presented in
Table~\ref{tab-TypeIIIKAB} for gases studied in this paper.  
The loss rate can also be due to photochemical dissociation of species $A$,
as in
\begin{equation}
\label{eq:lossrate2}
\label{eq:totallossrate}
L[A] = J[A] = \int_{\lambda} q_{\lambda} I_{\lambda} \exp^{-\tau_\lambda} \sigma_{\lambda}[A]d\lambda,
\end{equation}
where $J$ is the photodissociation loss rate, $q_{\lambda}$ is the
quantum yield, I$_{\lambda}$ is the stellar intensity,
$\exp^{-\tau_\lambda}$ is the attenuation by optical depth
$\tau_\lambda$, $\sigma$ is the photodissociation cross section of the
species $A$, and $\lambda$ is wavelength. Photodissociation is most
relevant high in the atmosphere typically above mbar levels to which
stellar UV radiation can penetrate from above.

\begin{table*}[ht]
\begin{center}
\begin{tabular}{|l|lll|lll|}
\tableline
\tableline
Reaction & $A$ & $n$ & $E$ & $T=270$ K & $T=370$ K & $T=470$ K \\ 
\tableline
{DMS + H $\rightarrow$ CH$_3$SH + CH$_3$ }  & $4.81\times10^{-18}$ & 1.70 & 9.00   & $2.63\times10^{-24}$ & $2.11\times10^{-22}$ & $2.91\times10^{-21}$ \\
{CH$_3$Cl + H $\rightarrow$ CH$_3$ + HCl }  & $1.83\times10^{-17}$ & 0 & 19.29  & $1.97\times10^{-24}$ & $1.46\times10^{-22}$ & $1.92\times10^{-21}$ \\
{CH$_3$Br + H $\rightarrow$ CH$_3$ + HBr }  & $8.49\times10^{-17}$ & 0 & 24.44  & $1.59\times10^{-21}$ & $3.01\times10^{-20}$ & $1.63\times10^{-19}$ \\
{CH$_3$I + H $\rightarrow$ CH$_3$ + HI }  & $2.74\times10^{-17}$ & 1.66 & 2.49  & $7.67\times10^{-18}$ & $1.75\times10^{-17}$ & $3.09\times10^{-17}$  \\
\hline
{DMS + OH $\rightarrow$ CH$_3$SCH$_2$ + H$_2$O }  & $1.13\times10^{-17}$ & 0 & 2.10   & $4.43\times10^{-18}$ & $5.71\times10^{-18}$ & $6.60\times10^{-18}$ \\
{CH$_3$Cl + OH $\rightarrow$ CH$_2$Cl + H$_2$O} & $1.40\times10^{-18}$ & 1.60 & 8.65  & $2.54\times10^{-20}$ & $1.89\times10^{-19}$ & $3.17\times10^{-19}$ \\
{CH$_3$Br + OH $\rightarrow$ CH$_2$Br + H$_2$O }  & $2.08\times10^{-19}$ &1.30 & 4.16  & $2.87\times10^{-20}$ & $7.13\times10^{-20}$ & $1.30\times10^{-19}$ \\
{CH$_3$I + OH $\rightarrow$ CH$_2$I + H$_2$O }  & $3.10\times10^{-18}$ & 0 & 9.31  & $4.90\times10^{-20}$ & $1.50\times10^{-19}$ & $2.86\times10^{-19}$  \\
\hline
{DMS + O $\rightarrow$ CH$_3$SO + CH$_3$ }  & $1.30\times10^{-17}$ & 0 & -3.40   & $5.91\times10^{-17}$ & $3.93\times10^{-17}$ & $3.10\times10^{-17}$ \\
{CH$_3$Cl + O $\rightarrow$ CH$_2$Cl + OH} & $1.74\times10^{-17}$ & 0 & 28.68  & $1.77\times10^{-23}$ & $8.77\times10^{-22}$ & $8.26\times10^{-21}$ \\
{CH$_3$Br + O $\rightarrow$ CH$_2$Br + OH }  & $2.21\times10^{-17}$ & 0 & 30.76  & $1.77\times10^{-23}$ & $8.77\times10^{-22}$ & $8.26\times10^{-21}$ \\
{CH$_3$I + O $\rightarrow$ CH$_3$ + IO }  & $6.19\times10^{-18}$ & 0 & -2.84  & $2.19\times10^{-17}$ & $1.56\times10^{-17}$ & $1.28\times10^{-17}$  \\
\tableline
\end{tabular}
\end{center}
\caption{Reaction rates with {H}, {OH}, and {O} of select Type-III biosignature gases. Second order reaction rates in units of m$^3$ molecule$^{-1}$ s$^{-1}$ are computed from the formula $k(T) = A (T/298)^n \exp(-E/RT)$ where $T$ is the temperature in K and $R$ is the gas constant ($R=8.314472\times10^{-3}$ kJ~mole$^{-1}$). The reactions rate are compiled from the NIST Chemical Kinetic Database. }
\label{tab-TypeIIIKAB}
\end{table*}


Gases can be lost from the atmosphere by deposition to the ground.
This loss is at the surface only (and numerically is treated as a
lower boundary condition), in contrast to the photochemical loss rate
reactions which take place throughout the upper atmosphere. Dry
deposition is deposition onto a surface (either solid land or liquid
water oceans) and wet deposition is deposition into water (rain)
droplets \citep{sein2006}.

The deposition velocity at the planetary surface can be described by
\begin{equation}
\label{eq:vdep}
\Phi_{\rm dep} = nv_d,
\end{equation}
where $\Phi$ is the molecular loss flux at the surface due to dry
deposition, $n$ is the number density (in units of molecules~m$^{-3}$)
at the surface of the species under consideration, and $v_d$ is the
dry deposition velocity (in units of m~s$^{-1}$). The wet deposition
is relevant for water-soluable gases and is not usually described as a
velocity but as a process that occurs througout the layer where water
condenses (on Earth the troposphere) \citep{hu2012}.  Where
photochemical reactions are slow, the deposition rate (the rate of
loss to the ground) can control the atmospheric concentration of
gas. Surface deposition consists of two processes: transfer of a gas
between the atmosphere and the surface and removal of the gas from the
surface.  The rate of transfer from the atmosphere to the surface is
proportional to the concentration difference between the atmosphere
and the surface. Thus once transferred to the surface, the gas has to
be chemically removed, or the surface will saturate with gas there
will be no more transfer---the deposition rate will be zero.  The
values of wet and dry deposition velocities can be measured on Earth
\citep[e.g.,][]{sehm1980}, but the wet and dry deposition rates for
various gases are highly variable and therefore averages tend to be
used in models. 

Caution must be taken, however, in applying Earth-based averages to
exoplanets. The chemistry that removes many of Earth's atmospheric
gases at the surface such as methane, ammonia, OCS and methyl chloride
from Earth's atmosphere are biochemical, not geochemical. Life
actively consumes these gases. So deposition rates on exoplanets may
be very different from those in the terrestrial atmosphere. We discuss
this in more detail in the context of CH$_4$ and NH$_3$ in
\S\ref{sec-methanogenesis} and \S\ref{sec-haber} respectively. In
summary, caution should be taken when extrapolating the Earth measured
values to planetary applications. Notably, if the surface is saturated
with a specific molecule, the surface uptake of the molecule may be
reduced to zero.

The production rate are written in terms of the two different loss
rates (and considering $\Phi_{\rm dep}$ as a surface boundary
condition,
\begin{equation}
\label{eq:productionrate}
P = [A](J + K_{AB}[B]).
\end{equation}
The production rate is here assumed
to be from biological sources, but when considering false positives
the geological source should also be considered. 

For biosignature gases that are minor chemical perturbers in the
atmosphere, the biosignature lifetime can be estimated based on the
dominant loss rate via equation~(\ref{eq:lossrate1}).  The simplified
example for one species $A$ gets more complicated for the case where
there are several terms in the loss rate and when the production rate
also includes other chemical reactions, and this is where the
photochemistry model calculation is required. The steady state
concentration $[B]$ is unknown and calculating $[B]$ is one reason why
a full photochemical model is needed to go beyond estimates.

The full photochemical model is presented in \citet{hu2012}. The
photochemical code computes a steady-state chemical composition of an
exoplanetary atmosphere. The system can be described by a set of
time-dependent continuity equations, one equation for each species at each
altitude. Each equation describes: chemical production; chemical loss;
eddy diffusion and molecular diffusion (contributing to production or
loss); sedimentation (for aerosols only); emission and dry deposition
at the lower boundary; and diffusion-limited atmospheric escape for
light species at the upper boundary. The code includes 111 species,
824 chemical reactions, and 71 photochemical reactions.

Starting from an initial state, the system is numerically evolved to
the steady state in which the number densities no longer
change. Because the removal timescales of different species are very
different, an implicit inverse Euler method is employed for numerical
time stepping. The generic model computes chemical and photochemical
reactions among all O, H, N, C, S species, and formation of sulfur and
sulfate aerosols. The numerical code is designed to have the
flexibility of choosing a subset of species and reactions in the
computation. The code therefore has the ability to treat both oxidized
and reduced conditions, by allowing selection of ``fast species''.  For
the chemical and photochemical reactions, we use the most up-to-date
reaction rate data from both the NIST database
(http://kinetics.nist.gov) and the JPL publication (Sander et
al. 2011). Ultraviolet and visible radiation in the atmosphere is
computed by the $\delta$-Eddington 2-stream method with molecular
absorption, Rayleigh scattering and aerosol Mie scattering
contributing to the optical depth.

We developed the photochemistry model from the ground up from basic
chemical and physical principles and using both established and
improved computer algorithms \citep[see][and references therein]{hu2012}.
We tested and validated the model by reproducing the atmospheric
composition of Earth and Mars. For one of the tests, we simulated
Earth's atmosphere starting from an 80\% N$_2$ and 20\%
O$_2$ atmosphere and temperature profile of the US Standard Atmosphere
1976. Important atmospheric minor species are emitted from the lower
boundary by standard amounts (e.g., reported by the IPCC), including
CO, CH$_4$, NH$_3$, N$_2$O, NO$_{\rm x}$, SO$_2$, OCS, H$_2$S,
H$_2$SO$_4$. We validate that vertical profiles predicted by our
photochemical model of O$_3$, N$_2$O, CH$_4$, H$_2$O, OH, HO$_2$, NO,
NO$_2$, HNO$_3$ match with various balloon and satellite observations,
and the surface mixing ratios of OH, O$_3$, SO$_2$ and H$_2$S also
match with standard tropospheric values \citep[see][for
  details]{hu2012}.  As another test, we reproduce the chemical
composition of the current Mars atmosphere, in agreement with measured
compositions \citep[e.g.,][]{kran2006} and previous 1-D photochemistry
model results \citep[e.g.,][]{zahn2008}. In particular, the code
correctly illustrates the effect of HO$_x$ catalytic chemistry that
stabilizes Mars' CO$_2$-dominated atmosphere, predicting an O$_2$
mixing ratio of $\sim$1500~ppb.

\subsection{Atmosphere Model}

We compute synthetic spectra of the modeled exoplanet's atmospheric
transmission and thermal emission with a line-by-line radiative
transfer code \citep{seag2000b, mill2009, madh2009}. Opacities are
based on molecular absorption with cross sections computed based from
the HITRAN 2008 database \citep{roth2009}, molecular collision-induced
absorption when necessary \citep[e.g.,][]{bory2002}, Rayleigh
scattering, and aerosol extinction computed based on Mie
theory. The transmission is computed for each wavelength by
integrating the optical depth along the limb path, as outlined in
\citep[e.g.,][]{seag2000a}. The planetary thermal emission is
computed by integrating the radiative transfer equation without
scattering for each wavelength \citep[e.g.,][]{seag2010}.

The temperature profiles for the simulated atmospheres are
self-consistently computed with the grey atmosphere assumption
\citep{guil2010}. Opacities of CO$_2$, H$_2$O, CH$_4$, O$_2$, O$_3$,
OH, CH$_2$O, H$_2$O$_2$, HO$_2$, H$_2$S, SO$_2$, CO, NH$_3$ are
considered, if they are needed in the atmospheric scenario under
consideration. For the grey-atmosphere temperature profiles, we have
assumed isotropic irradiation, and applied the convection correction
if the radiative temperature gradient is steeper than the adiabatic
lapse rate.  We have assumed for all cases the planetary Bond albedo
is 0.3; in other words, we have implicitly assumed a cloud coverage of
50\% (assuming cloud albedo to be 0.6). For consistency we account for
the effect of clouds on the planetary reflection and thermal emission
spectrum by a weighted average of spectra with and without clouds as
in \citet{desm2002}.

We emphasize that the precise temperature-pressure structure of the
atmosphere is less important than photochemistry for a first-order
description of biosignature gases.

\subsection{Detection Metric}
\label{sec-detectionmetric}
We now describe our metric for a ``detection'' that leads to a
required biosignature gas concentration.  For now, the detection has
to be a theoretical exercise using synthetic data. We determine the
required biosignature gas concentration based on a spectral feature
detection with a SNR=10.  Specifically, we describe the SNR of the
spectral feature as the difference between the flux in the absorption
feature and the flux in the surrounding continuum (on either side of
the feature) taking into account the uncertainties on the data,
\begin{equation}
{\rm SNR} = \frac{\left| F_{out} - F_{in} \right|}{\sqrt{\sigma^2_{F_{out}} +
  \sigma^2_{F_{in}}}},
\end{equation}
where $F_{in} \pm \sigma_{F_{in}}$ is the flux density inside the
absorption feature and $F_{out} \pm \sigma_{F_{out}}$ is the flux
density in the surrounding continuum, and $\sigma$ is the uncertainty
on the measurement.

The uncertainties of the in-feature flux and continuum flux are
calculated for a limiting scenario: an Earth-sized planet orbiting a
noiseless Sun-like star at 10 pc observed (via direct imaging) with a
6 m-diameter telescope mirror operating within 50\% of the shot noise
limit and a quantum efficiency of 20\%. The integration time is
assumed to be 20 hours. We note that collecting area, observational
integration time, and source distance are interchangeable depending on
the time-dependent observational systematics.

\section{Results}
\label{sec-results}

Our results are the biomass estimates for exoplanet atmosphere
scenarios with select biosignature gases. We present a few case
studies, including both familiar biosignature and bioindicator gases
and biosignature gases not yet widely discussed. The illustrative
examples are for thermal emission spectra. A later paper also treats
transmission spectra \citep{seag2013}. The case studies aim to
demonstrate the use of the biomass model: to take a biosignature gas,
let the source flux be a free parameter (instead of tied to Earth-life
production rates), and check that the biomass is physically plausible.

\subsection{NH$_3$ as a Biosignature Gas in a Reducing Atmosphere}
\label{sec-haber}

We propose ammonia, NH$_3$, as a biosignature gas in an H$_2$-rich
exoplanet atmosphere.  NH$_3$ is a good biosignature gas candidate for
any thin H$_2$-rich exoplanet atmosphere because of its short lifetime
and lack of production sources. NH$_3$ as a biosignature gas is a new
idea, and one that is specific to a non-Earth-like planet. On Earth,
NH$_3$ is not a useful biosignature gas because, as a highly valuable
molecule for life that is produced in only small quantities, it is
rapidly depleted by life and unable to accumulate in the atmosphere.

The biosignature idea is that ammonia would be produced from hydrogen
and nitrogen, in an atmosphere rich in both,
\begin{equation}
\label{eq:haber}
{\rm 3H_2 + N_2 \rightarrow 2 NH_3}.
\end{equation}
This is an exothermic reaction which could be used to capture energy.
We propose that in an H$_2$-rich atmosphere, life can catalyze
breaking of the N$_2$ triple bond and the H$_2$ bond to produce
NH$_3$, and capture the energy released. 
In an H$_2$-rich environment, life could use the reduction of
N$_2$ to capture energy---in contrast
to life on Earth which solely fixes nitrogen in an energy-requiring process.
Energy capture would yield an excess of ammonia over
that needed by life to build biomass, and so the excess would
accumulate in the atmosphere as a Type I biosignature gas.

A catalyst is required to synthesize ammonia from hydrogen and
nitrogen gas because the reaction in equation~(\ref{eq:haber}) does
not occur spontaneously at temperatures below 1300~K, at which
temperature the formation of ammonia is strongly thermodynamically
disfavored.  On Earth, the industrial production of NH$_3$ by the
above reaction is called the Haber process: an iron catalyst is used
at high pressure (150-250 bar) to allow the reaction to happen at
575-825~K at which temperature the formation of ammonia is
thermodynamically favored \citep{habe1913}.  The Haber process is the
principle industrial method for producing ammonia\footnote{Fritz Haber
  and Carl Bosch were awarded the Nobel Prize for this chemistry in
  1918, ironically, as the Haber process' principal deployment in the
  previous four years had been for making explosives for munitions,
  the application of chemistry that Alfred Nobel most wanted to
  discourage.}.  More efficient catalysts are known, which can
catalyze the formation of ammonia from elemental nitrogen and hydrogen
gases at 500~K and standard pressure \citep{yue2006,avan2007}. Others
catalyze the formation of ammonia from nitrogen gas and a proton
\citep{yand2003, shil2003, wear2006} or from an activated nitrogen
molecule and hydrogen gas \citep{nish1998}, in water at room
temperature and pressure. The final step of combining a room
temperature nitrogen reduction catalyst with a room temperature
hydrogen splitting catalyst has not been achieved in the lab but is
believed to be a realistic goal \citep{wear2006}. Such a combined
catalyst would make NH$_3$ from H$_2$ and N$_2$ and would generate
energy (heat) from the reaction.

Life on Earth does not use the ``Haber'' reaction. This might be
because the appropriate catalysts have not evolved. It might be
because the rare environments in which H$_2$ is available provide
other more easily accessible sources of energy (such as
methanogenesis; \S\ref{sec-methanogenesis}). On Earth life inputs
energy to break the N$_2$ triple bond, and the fixed nitrogen is a
valuable resource representing an investment of energy and so is
avidly taken up by other Earth life. Haber life using the Haber
chemistry in an atmosphere with plenty of N$_2$ would produce generous
amounts of NH$_3$, more than enough for the rest of life to use,
enabling NH$_3$ to accumulate in the atmosphere as a biosignature gas.

\begin{figure*}[ht]
\begin{center}
\includegraphics[scale=.55]{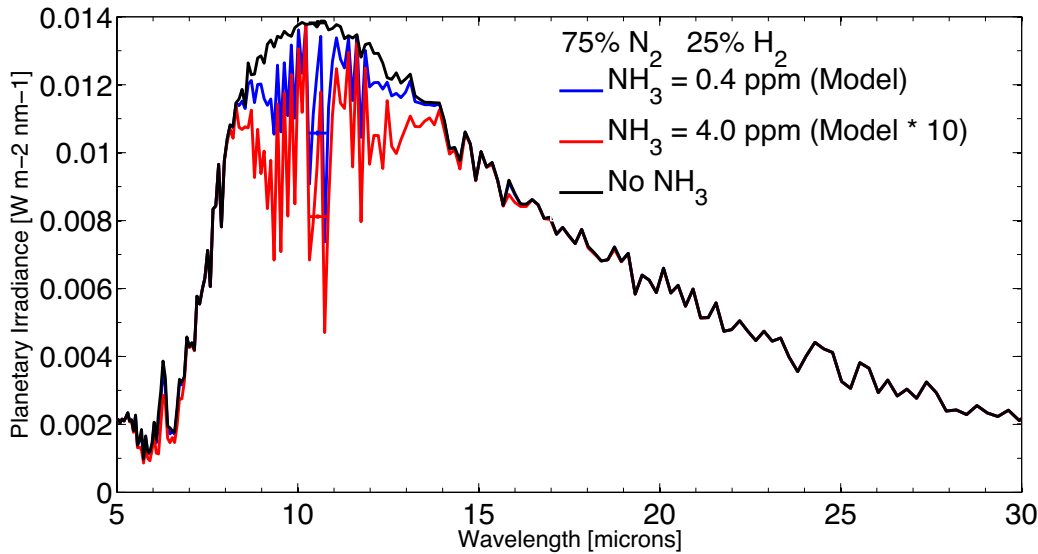}
\caption{Synthetic thermal emission spectra for the ``cold Haber
  World''.  The 1 bar atmosphere is composed of 75\% N$_2$ and 25\%
  H$_2$, with a 290~K surface temperature.  NH$_3$ (that would be
  produced by life), is emitted from the surface.  The spectrum in
  blue is computed from atmospheric composition calculated by the
  photochemistry model (the blue line in this Figure corresponds
  directly to the composition shown in
  Figure~\ref{fig:NH3mixingratios}).  The spectra in red and black are computed with no NH$_3$
  and 10 times more NH$_3$, respectively, for comparison. The spectra
  are computed at high spectral resolution and binned to a spectral
  resolution of 100. The horizontal bars show the broadband flux in
  the 10.3-10.8 micron band, most sensitive to the atmospheric NH$_3$
  feature.}
\label{fig:NH3spectrum}
\end{center}
\end{figure*}

\begin{figure*}[ht]
\centering
\includegraphics[scale=.75]{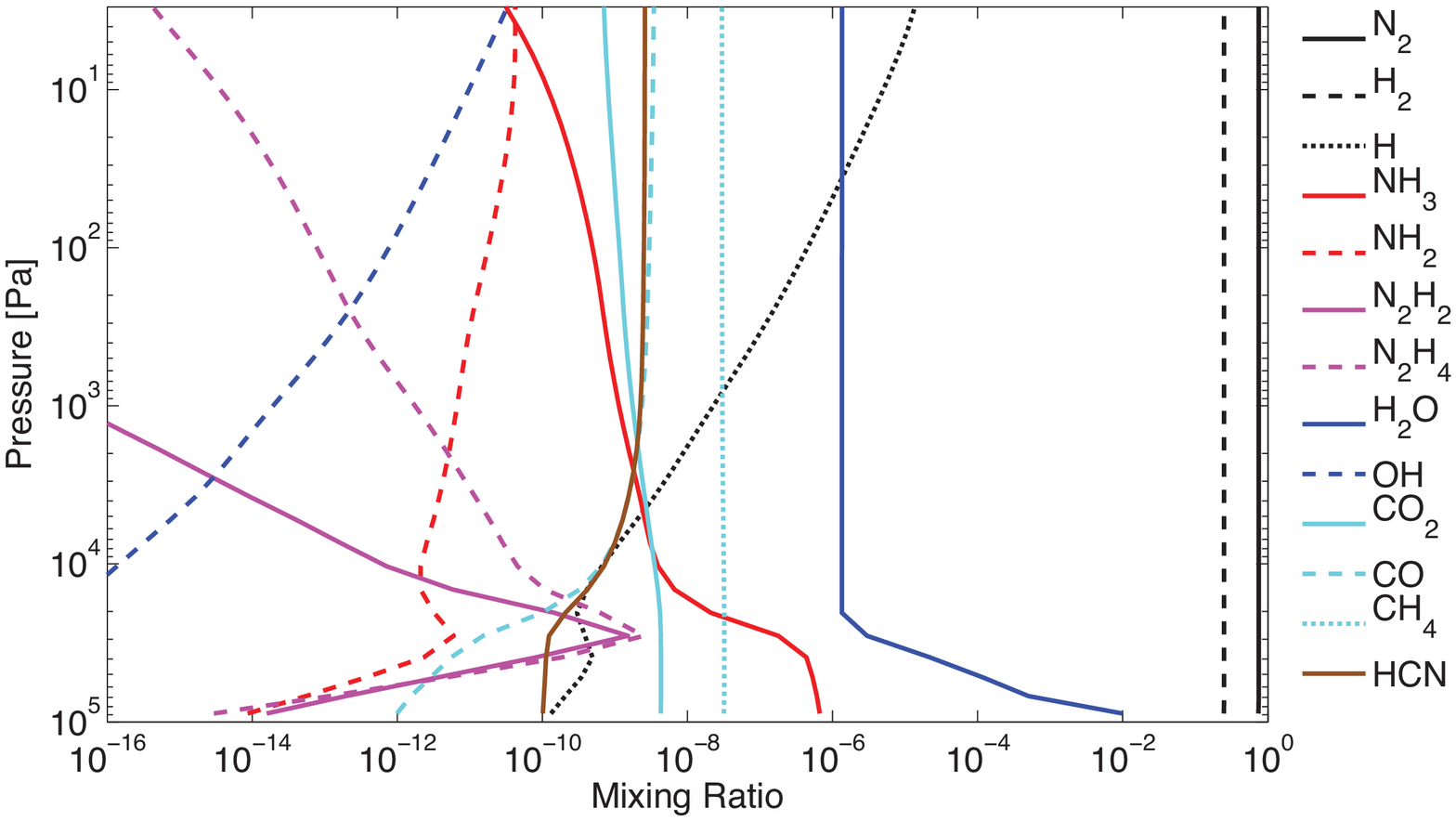}
\caption{Mixing ratios for the atmosphere of the ``cold Haber World''.
  The 1 bar atmosphere is composed of 75\% N$_2$ and 25\% H$_2$, with
  a 290~K surface temperature.  The simulated planet is an Earth-sized
  planet at 1.1 AU from a Sun-like star. The planet's surface is a net
  emitter of NH$_3$, with a surface source flux of $5 \times
  10^{15}$~molecule~m$^{-2}$~s$^{-1}$ computed to match the 0.1 ppm
  concentration required by our detection metric. (Note that the
  present-day Earth's ammonia emission rate is $8 \times
  10^{13}$~molecule~m$^{-2}$~s$^{-1}$.) We also include CO$_2$
  emission of $1\times10^{14}$~molecule~m$^{-2}$~s$^{-1}$ (1 order of
  magnitude smaller than Earth's volcanic CO$_2$ emission). Our
  photochemistry model shows that NH$_3$ can accumulate to a mixing
  ratio of 0.4 ppm in the convective layer of the atmosphere while it
  is destroyed in the upper atmosphere layers.
}
\label{fig:NH3mixingratios}
\end{figure*}

To check the viability of NH$_3$ as a biosignature gas, we follow the
steps listed in \S\ref{sec-biomassmodels}.  For background, we
consider a planet of Earth mass and size, a 290~K surface temperature
and with a 1-bar atmosphere composed of 25\% H$_2$ and 75\% N$_2$ by
volume, and including carbon species via a CO$_2$ emission flux from
the planet's surface.  NH$_3$ has significant opacity in the 10.3-10.8
$\mu$m band in a thermal emission spectrum. A mixing ratio of 0.1 ppm
is radiatively significant for a 1-bar atmosphere in this band in an
H$_2$-N$_2$ atmosphere (Figure~\ref{fig:NH3spectrum}) according to our
detection metric (\S\ref{sec-detectionmetric}).  We next determine the
NH$_3$ surface source flux required for the gas to accumulate in the
atmosphere to the 0.1 ppm concentration level. For this, we compute
the photochemical equilibrium steady-state composition (results shown
in Figure~\ref{fig:NH3mixingratios}) with the ammonia surface source
flux as a free parameter (\S\ref{sec-photochemmodel}). The dominant
loss mechanism of NH$_3$ is due to photolysis (or reaction with OH;
each process breaks the NH$_3$ bond) with some NH$_3$ eventually being
converted to N$_2$).  We adopt an NH$_3$ deposition velocity of
0~m~s$^{-1}$, assuming that the surface is saturated in NH$_3$.  (See
the below for a further discussion of deposition removal rate
assumptions and related consequences).  The resulting NH$_3$ surface
source flux is $5.0 \times 10^{15}$~molecule~m$^{-2}$~s$^{-1}$.  We
note that this planet has a column-averaged mixing ratio of 0.4 ppm of
NH$_3$, and to meet a surface temperature of 290~K the semi-major axis
would be 1.1~AU. We also point out that NH$_3$ concentrates mostly in
the lower atmosphere, and decreases very rapidly with altitude above
15 km (Figure~\ref{fig:NH3mixingratios}) because of the high-altitude
destruction rates. To compute $\Delta G$ we used $T=290$~K, and
reactant and product concentrations at the surface in terms of partial
pressures of N$_2$ = 0.75, H$_2$ = 0.25, NH$_3 = 6.6 \times 10^{-7}$.

We next estimate the biomass surface density. Using the biomass
equation (equation~(\ref{eq:TypeIBiomassModel})) with the NH$_3$
source flux of $5.0 \times 10^{15}$~molecule~m$^{-2}$~s$^{-1}$ (or
$8.4 \times 10^{-9}$~mole~m$^{-2}$s$^{-1}$); $\Delta G =
75.0$~kJ~mole$^{-1}$ at 300~K;
and $P_{m_e} = 7.0 \times 10^{-6}$ kJ~g$^{-1}$~s$^{-1}$, we find a
biomass surface density of $8.9 \times 10^{-2}$~g~m$^{-2}$.  We
therefore consider the NH$_3$ production flux to be viable in our
Haber World scenario.  The global annual biogenic NH$_3$ surface
emission in the Haber World would be about 6700~Tg~yr$^{-1}$.  This is
much higher than the Earth's natural NH$_3$ emission at
10~Tg~yr$^{-1}$ \citep{sein2006}. Comparing NH$_3$ production on the
Haber world and on Earth, however, is not valid. We are postulating
that production of NH$_3$ on the Haber world is a major source of
metabolic energy for life. A better emission rate comparison is to the
biosignature gas O$_2$ from Earth's principle energy metabolism,
photosynthesis. Earth's global oxygen flux is 500 times larger than
the Haber World's NH$_3$ surface emission, at about $2 \times
10^{5}$~Tg~yr$^{-1}$ \citep{frie2009}.

Deposition rates require more description, because on Earth deposition
is the dominant atmospheric removal process for NH$_3$, yet we argue
the deposition rates to play a minor role for NH$_3$ atmospheric
removal on the Haber World. On Earth, ammonia is taken up avidly by
life in soil and in water.  On our proposed Haber World, life is a net
producer of ammonia, not a net consumer, and so water (as raindrops or
ocean) and soil would saturate with ammonia. Wet and dry deposition
rates would therefore be limited to chemical consumption, the rates of
which would depend on specifics of the surface chemistry. We therefore
assume that the deposition rates are much less than atmospheric loss
rates and this is considered in the photochemical calculation. A
larger deposition rate implies a larger NH$_3$ emission flux and
therefore a larger biomass to maintain the biosignature. When
deposition is the dominant removal process, the relation between the
source flux and the steady-state mixing ratio of NH$_3$ is linear.  A
Haber World with Earth-like NH$_3$ deposition rates
($10^{-3}$~m~s$^{-1}$)requires a NH$_3$ source flux of about $3.0
\times 10^{17}$~molecule~m$^{-2}$~s$^{-1}$, a surface emission flux
100 times higher than the zero-deposition rate case. The biomass is
also 100 times higher, and at $\Sigma_B \sim 9$~g~m$^{-2}$ is still a
reasonable value.

The molecule HCN can be considered a bioindicator gas in the specific
situation where HCN is detectable but its formation components,
NH$_3$ and CH$_4$ (or any other dominant carbon source such as CO or
CO$_2$) are not. More specifically, in an H$_2$-rich atmosphere, if
NH$_3$ and CH$_4$ are emitted at comparable rates, HCN will be
produced with little NH$_3$ and CH$_4$ remaining in the atmosphere
above the convective layer \citep[or above the first several scale
  heights from the surface;][]{hu2012}. It can therefore be expected
that if emissions of methane and ammonia are both elevated, the mixing
ratio of HCN in the atmosphere can be as high as 1 ppmv and then
become detectable via its prominent spectral feature at $\sim$3~microns.
In other words, HCN can be an indicator of surface NH$_3$ emission,
even if NH$_3$ itself is depleted and not detectable due to
atmospheric photochemistry.  The photochemical pathway of HCN under
such conditions is described in detail in \citet{hu2012}.  In
general, in an atmosphere with enough NH$_3$ or N$_2$ and CH$_4$, the
formation of HCN is inevitable in anoxic environments
\citep{zahn1986}. Also of potential interest, HCN is the second most
common N-bearing species in the Haber World.

In terms of false positives for NH$_3$, the NH$_3$ biosignature gas
concept is not changed in the massive atmosphere case with high
surface pressure. As long as the surface conditions are suitable for
liquid water, NH$_3$ will not be created by uncatalyzed chemical
reactions. False postives may still exist such as chemical or
biological breakdown of abiotic molecules. An additional false
positive for NH$_3$ could be generated by non-life-compatible surface
temperatures: at 820~K with surface iron, NH$_3$ could be generated by
the conventional Haber process. These false positive statements hold
for a rocky planet with a thin atmosphere; other cases such as planets
with a massive atmosphere where NH$_3$ could be generated kinetically
at extremely high pressures, or planets withh icy interiors where
NH$_3$ is outgassed during planetary evolution, have to be assessed on
a case-by-case basis (see \citet{seag2013}).

In summary, we propose NH$_3$ as a biosignature gas on a planet with
an N$_2$-H$_2$ composition.  NH$_3$ should be photodissociated and
therefore its presence would be indicative of biogenic production. We
have nicknamed this planet ``cold Haber World'' because life would
have to perform the Haber process chemistry using a highly efficient
catalyst, low temperature to break both the N$_2$ triple bond and the H$_2$
bond, rather than the elevated temperature and relatively inefficient
catalyst used in the industrial ``hot'' Haber process.

\subsection{CH$_4$ on Terrestrial-Like Exoplanets}
\label{sec-methanogenesis}

We revisit methane as a biosignature gas on early Earth, to estimate
the biomass surface density required for primary producers to generate
a remotely detectable CH$_4$ concentration.

CH$_4$ has long been considered a prime biosignature gas for
Earth-like exoplanets \citep{hitc1967} and especially for early Earth
analogs \citep{desm2002}. An early Earth analog prevalence of CH$_4$
theory is motivated by the early faint young sun paradox. A few
billion years ago, the sun was 20-30\% fainter than today,
specifically with 26\% lower luminosity 4 Gyr ago, based on
asteroseismology-constrained stellar evolution models
\citep{bahc2001}. Yet, there is no evidence that Earth was frozen over
during that time. A reduced greenhouse gas is a good, viable
explanation for keeping Earth warmed despite the much cooler sun
\citep{saga1972}. The greenhouse gas CH$_4$ is a favored greenhouse
gas explanation \citep{kieh1987, haqq2008}. Methane at 1,000 times
today's atmospheric concentration would have been sufficient to keep
the Earth warm (i.e., concentrations of 1600 ppmv instead of 1.6 ppmv)
\citep[][and references therein]{pavl2000, haqq2008}. Moreover,
accumulation of atmospheric CH$_4$ to levels much higher than Earth's
would have been possible in the anoxic environment of early Earth.
CH$_4$ could have accumulated to 1000 ppmv levels in the weakly
reducing early Earth atmosphere because the CH$_4$ loss rate was so
much smaller, owing to the lack of O$_2$.  In more detail, the
dominant removal rate of methane in an oxidized atmosphere is due to
reaction with atmospheric OH. OH is produced via photochemistry, from
both O$_3$ (itself a photochemical byproduct of O$_2$) and H$_2$O
\citep{sein2006}.

To consider the biomass required on any methanogenesis world, we must
work with surface fluxes generated only by primary production
methanogenesis.  Overlooked or unstated for early Earth in past work
is that most of Earth's biogenic methane production today is from
fermentation of biomass. This biomass available 
for fermentation is almost entirely produced via
photosynthesis.  For an early Earth analog, before the rise of
oxygenic photosynthesis, there is likely no large reservoir of biomass
for fermentation.  We note that on early Earth itself, there should
have been a small reservoir of biomass, from, for example, anoxic
photosynthesis.  For exoplanets we want to consider
biologically-produced methane from an ecosystem that uses
methanogenesis as a primary energy source.  In such a methanogenesis
world, there would be biomass for fermentation, but the amount of
methane produced by fermentation would be minor compared to the amount
of methane produced from energy capture.

The methanogens of interest are those that convert H$_2$ and CO$_2$ to
CH$_4$ in the process of extracting energy from the environment.
These methanogens do not require biomass to feed upon and today live in
anoxic environments including hydrothermal vents at the deep sea
floor, subterranean environments, and hot springs.

The methanogens of interest produce CH$_4$ by using carbon from CO$_2$
from inorganic sources,
\begin{equation}
\label{eq:methanogenesis}
{\rm H_2 + CO_2 \rightarrow CH_4 + H_2O}.
\end{equation}
At the deep-sea floor, H$_2$ is released from rocks by hot water
emitted from hydrothermal vents (serpentinization). CO$_2$ is
available dissolved in seawater. In other Earth environments, H$_2$ is
also produced as a byproduct of biological metabolism, and CO$_2$ is
available as gas in air or dissolved in water. The metabolic
byproducts from methanogenesis are CH$_4$ and H$_2$O.

\begin{figure*}[ht]
\begin{center}
\includegraphics[scale=.55]{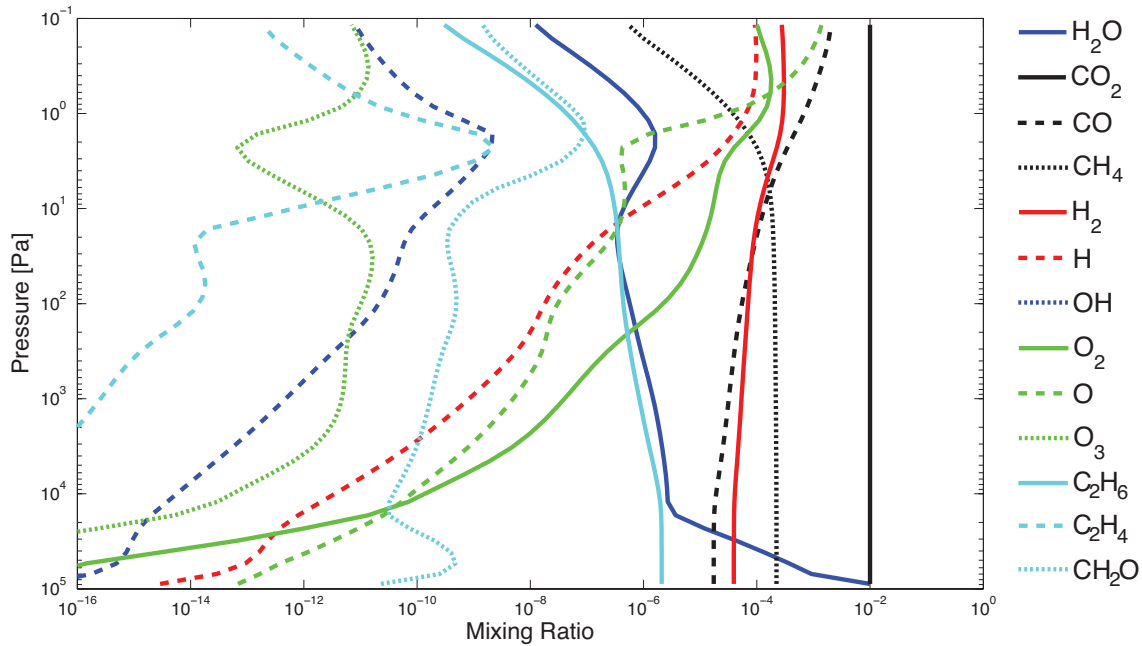}
\caption{Atmospheric composition for the early Earth methanogenesis
  world. The 1 bar atmosphere is taken to be N$_2$-dominated with 1\%
  CO$_2$. We take the planet's surface to be a net emitter
  of CH$_4$, with a surface flux of $7\times10^{14}$
  molecules~m$^{-2}$s$^{-1}$ computed to match the 200~ppm
  concentration required by our detection metric. Our photochemistry
  model shows that methane can accumulate to a mixing ratio of 220 ppm
  in the atmosphere and that the major photochemical product is H$_2$
  having a mixing ratio of 40~ppm and CO having a mixing ratio of
  18~ppm.}
\label{fig:CH4composition}
\end{center}
\end{figure*}

We now turn to the biomass estimate for CH$_4$ as the biosignature
from a primary producing methanogenesis ecology on an early Earth
analog exoplanet. We take early Earth (Earth-size, Earth-mass with a
1~bar atmosphere) to be an anoxic, N$_2$-dominated atmosphere with
CO$_2$ mixing ratios of 1\% and 20\% \citep{ohmo2004}, the time period
before the rise of oxygen (an Archaen atmosphere, 3.8 to 2.5 Gyr
ago). For the planets to have surface temperatures of 290~K, they
would be at 1.2 AU for the 1\%~CO$_2$ atmosphere and 1.3~AU for the
20\% CO$_2$ atmosphere.  Using our detection metric
(\S\ref{sec-detectionmetric}) We find a detectable CH$_4$ mixing ratio
to be 200 ppm in the 3.1 to 4 $\mu$m band. See
Figure~\ref{fig:CH4composition} for the chemical composition of the
1\%~CO$_2$ atmosphere. 

Although we consider the effect of clouds on weakening the spectral
features in the exoplanet spectrum, we do not treat formation of
hydrocarbons that have more than two carbon atoms. The formation of
hydrocarbons that have more than two carbon atoms should only have
minor impact on our estimate of required biomass, because the most
abundant hydrocarbons that have more than two carbon atoms,
C$_3$H$_8$, has still five orders of magnitude lower mixing ratio at
the steady state than C$_2$H$_6$ in N$_2$ atmospheres \citep{pavl2001}.
The formation of hydrocarbon haze (commonly hydrocarbons
with more than 5 carbon atoms), a point we do not
address in this paper, may impact the energy budget of the planet and
therefore the surface temperature \citep{haqq2008}.

A CH$_4$ source flux of 
$7.0 \times 10^{14}$~molecules~m$^{-2}$~s$^{-1}$ 
($1.2 \times 10^{-9}$~mole~m$^{-2}$~s$^{-1}$) 
and
$5.0 \times 10^{14}$~molecules~m$^{-2}$~s$^{-1}$ 
($8.3 \times 10^{-10}$~mole~m$^{-2}$~s$^{-1}$) 
is required to reach the 200 ppm CH$_4$
concentration in the 1\% and 20\% CO$_2$ atmospheres respectively.
To compute the $\Delta G$ value we used the
reactant and product molecule concentrations at the surface in terms
of partial pressures as follows. For the CO$_2$ case of 0.01 we used
H$_2 = 3.9 \times 10^{-5}$, CH$_4 = 2.2 \times 10^{-4}$, H$_2$O = 0.01, and
this results in a $\Delta G = 47.4$~kJ~mole$^{-1}$. In the CO$_2$ of
0.20 case we used H$_2 = 1 \times 10^{-5}$, CH$_4 = 2.5 \times 10^{-4}$,
and H$_2$O = 0.01, and this results in a $\Delta G =
41.2$~kJ~mole$^{-1}$.

Using the above values, together with $P_{m_e} = 7.0 \times
10^{-6}$~kJ~g$^{-1}$~s$^{-1}$ for the minimal maintenance energy rate
for anaerobic microbes at 290~K, the biomass estimate is $7.8 \times
10^{-3}$~g~m$^{-3}$ and $4.9 \times 10^{-3}$~g~m$^{-3}$, for the 1\%
and 20\% CO$_2$ atmospheres respectively, globally averaged values.
This is a reasonable biomass as compared to terrestrial microbial
biomass surface density values (Section~\ref{sec-biomassdensity}).  In
this scenario, methanogenic organisms would dominate a biosphere where
methanogenesis is the main energy source for life, just as on Earth
photosynthetic organisms (including plants and water-based
photosynthesizes) dominate the biosphere because photosynthesis is the
dominant energy source for life.

For an early Earth analog ``slime world'', the critical question is
whether H$_2$ would be a limiting input for methanogenesis on a planet
with Earth-like atmospheric conditions (see
equation~(\ref{eq:methanogenesis})).  For the methanogenesis world, a
global CH$_4$ flux of about 750~Tg~yr$^{-1}$ is required to reach the
detectability threshold (this global CH$_4$ flux is the value corresopnding to the CH$_4$ production rate calculated above).
To investigate we look at fluxes
of hydrogen gas from Earth's hydrothermal systems and levels of
hydrogen gas in hydrothermal fluids. Fluxes of hydrogen from
hydrothermal systems on Earth can be as high as 1.8\% of the total
gases \citep[see, e.g.,][]{gerl1980, legu1982, tara1991}. If this flux
of H$_2$ is extrapolated to the global fluxes by comparison with
fluxes of H$_2$S and SO$_2$ \citep{halm2002}, it would imply global
H$_2$ fluxes of between 0.8 and 1.6 Tg/year\footnote{On Earth we do
  not see this flux of H$_2$ into the atmosphere because most H$_2$ is
  consumed at the point of emission by microorganisms in
  methanogenesis, reduction of sulfate, or by direct oxidation with
  atmospheric oxygen, so little escapes to the atmosphere.}. If all of
the hydrogen were used in methanogenesis, that would result in 3.2-6.4
Tg~yr$^{-1}$ methane.

If the H$_2$ needed to support the required rate of methanogenesis
comes from volcanism alone---as in the case of modern Earth---a
methanogenesis world with an Earth-like atmosphere, requires about 100
times more hydrogen flux outgassing than on present day Earth.
This high H$_2$ flux outgassing could be sustained by either a more
reduced mantle \citep{holl1984, kast1993}, more serpentization, and/or
more volcanism than on the present-day Earth.

The required large amounts of H$_2$ could be also produced from
atmospheric photochemistry, in the absence of atmospheric oxygen,
enabling a biochemical cycle to sustain surface methanogenesis.  In
our photochemistry code, atmospheric H$_2$ forms from photolysis of
CH$_4$. The net accumulation of CH$_4$ results from the difference in
production and loss of CH$_4$, being computed self-consistenly with
the appropriate mass balance. A form of this methanogenesis scenario
was describe previously in \citet{khar2005}.  In this biochemical
cycle, a reasonable quantity of H$_2$ is required only at the onset of
the evolution of methanogenesis.

We note that methane can reach high abundances through abiotic means
especially if the mantle is reduced. More work is needed on false
positives in the hope of finding a way to distinguish biotic and
abiotic methane, or at least to assign a probability to the chance of
methane being biotic.

We finish the early Earth methanogenesis scenario with an emphasis on
the deposition velocity as a loss rate. For CH$_4$, we have considered
a zero deposition velocity, with the concept that a slime world
covered in methanogens has a CH$_4$-saturated surface. We also assume
that this world has few CH$_4$-consuming organisms compared to the
CH$_4$-producing organisms. In other words, the rationale is that on a
methane slime world, living organisms are net producers of CH$_4$ at
the surface and would not destroy CH$_4$---the surface would likely be
saturated in CH$_4$ making the deposition rate zero. On Earth CH$_4$
deposited to the surface is rapidly oxidized by life to CO$_2$, so
deposition is efficient, although only a minor contribution to the
CH$_4$ loss rate.  If we treat the N$_2$-CO$_2$ atmosphere with an
Earth-like CH$_4$ deposition rate (for CH$_4$ on Earth typically
$10^{-6}$~m~s$^{-1}$), then CH$_4$ is prevented from accumulating to
200 ppm, but instead is at a lower concentration, at values of 15~ppm. A
deposition velocity could be non-zero due to life other than the
methanogens consuming CH$_4$; but see the discussion in
\S\ref{sec-orderofmag}.

To summarize this subsection, we have revisited methane as a
biosignature gas on early Earth. We have found the biomass surface
density needed to sustain a detectable CH$_4$ biosignature gas
from primary production is reasonable at $\sim 5 \times 10^{-3}$~g~m$^{-2}$.
Although volcanism alone is unlikely to provide
the amount of H$_2$ needed to sustain methanogenesis at the level
required for CH$_4$ detection, an atmospheric photolysis of CH$_4$
can recycle the H$_2$ to provide sufficient flux.

\subsection{Martian Atmospheric Methane} 
\label{sec-Mars}

As a second case study we apply our biomass model to the methane flux
on Mars. CH$_4$ has been detected in the atmosphere of Mars with three
independent instruments \citep{form2004, kras2004, mumm2009}.  The
Martian CH$_4$ detection is difficult to reconcile with present
understanding of the planet and some believe the ground-based CH$_4$
detection may be a result of observational artifacts \citep[see the
  references in the summary review by][] {atre2011}. The Martian
CH$_4$ may be the result of geochemical outgassing or atmospheric
photochemistry (Bar-Nun and Dimitrov 2006). A more intriguing, if
speculative CH$_4$ source, is Martian life \citep{kras2004}. In this
subsection we show that the minimum required biomass density is
plausible for the measured CH$_4$ fluxes.

Averaged CH$_4$ levels are 5-30 ppbv in the spring and summer
mid-latitudes on Mars, depending on location, local time of day, and
season \citep{form2004, gemi2008, mumm2009, gemi2011}. There are
pronounced local hotspots for methane concentration, hence presumably
for CH$_4$ production. Different CH$_4$ observational studies,
however, find hotspots in different regions \citep[e.g.,
  compare][]{mumm2009, gemi2011}, so we use global averages for our
case study. The photochemical loss rate for the 5-30 ppb average level
of CH$_4$ is 1~to~2~$\times 10^9$ molecules~m$^{-2}$~s$^{-1}$
\citep{wong2004, kras2004}

If Martian organisms were producing CH$_4$, they would be reducing
atmospheric CO$_2$ with mantle-derived material, most plausibly H$_2$,
so that the methane we observe would be the product of
methanogenesis. We can calculate the free energy available from the
reaction
\begin{equation}
\label{eq:marsmethanogenesis}
{\rm 4H_2 + CO_2 \rightarrow CH_4 + 2H_2O},
\end{equation}
and hence estimate the biomass present in the soil that might account
for the observed methane flux.  The Gibbs free energy $\Delta G$ for
methanogenesis under Martian daytime maximum temperature range
(250-265K) is calculated assuming the 15~ppb CH$_4$ surface flux, a
likely surface hydrogen concentration of 15~ppm \citep{kras2001}, a
CO$_2$ partial pressure of 0.95, an an H$_2$O partial pressure of
$3 \times 10^{-4}$. $\Delta G$ is 85.2 to 77.5~kJ~mole$^{-1}$.  $P_{m_e} \simeq
7 \times 10^{-8}$ and $\simeq 5 \times 10^{-7}$
for temperatures at 250~K and 265~K respectively. Combined
with the surface flux above the CH$_4$ hotspots of the 1~to~2~$\times
10^9$ molecules~m$^{-2}$~s$^{-1}$ \citep{wong2004, kras2004}, we find
from equation~(\ref{eq:bio3}) surface biomass density of $\Sigma_B
\simeq 10^{-6}$--$10^{-7}$. This a very small biomass surface density
as compared to terrestrial biofilm values
(\S\ref{sec-biomassdensity}), and hence the Martian CH$_4$ production
by microbial life appears physically plausible.

The Martian surface is believed to be sterile, in part because the
surface atmospheric pressure is incompatible with the existence of
liquid water at any temperature and in part because the surface is
unshielded from extremely destructive radiation from space (Solar UV
and X rays, cosmic rays) (Dartnell 2011). Water, if it exists near the
surface, will be present as ice. Viking's finding of a complete lack
of organic molecules in the top few centimeters of soil supports the
sterility of the surface \citep{biem1976, biem1977}.  A few tens of
centimeters of regolith would shield organisms from radiation
\citep{pavl2002}. However orbital radar suggests that water is frozen
to a depth of several km in most sites on Mars (reviewed in Kerr
2010), so Martian life must either be more deeply buried than current
radar penetration, or be living in highly concentrated brines at
depths of 1 to 3 km.  The column-integrated density of $\sim
10^{-6}$--$10^{-7}$~g~m$^{-2}$ of biomass would therefore be present
not as surface life but living in rock interstices. This density is
well below that of the density of such rock-dwelling microbial
communities on Earth \citep[][and references therein]{pede1993}.

In summary, we have applied our biomass model to the putative methane
detections on Mars. We found that if the CH$_4$ is produced by
methanogenesis only a very small biomass surface density is required,
$\Sigma_B \simeq 10^{-6}$--$10^{-7}$~g~m$^{-2}$.  Martian methane
could be generated by microorganisms living in subsurface rocks. Our
model predicts that the amount of biomass needed to generate the
proposed methane flux is plausible for a rock-based microbial
community.  By itself, a biomass surface density prediction does not
rule out methanogenesis as the cause of the atmospheric CH$_4$.

\subsection{H$_2$S: An Unlikely Biosignature Gas}
\label{sec-H2S}

The gas H$_2$S is generated by bacteria on Earth, and also by
volcanism. The majority of Earth's H$_2$S emission is from life
\citep{watt2000}.  With our biosignature framework we can calculate a
consistent biosphere, based on sulfur-metabolizing organisms and
estimate how much biomass is required to generate a detectable amount
of H$_2$S. Although detection of H$_2$S will be very difficult due to
water vapor contamination of H$_2$S features, and just as seriously
H$_2$S has a serious risk of false positive through volcanic
production, the biomass estimate turns out to be reasonable.

H$_2$S would be very difficult to detect in a future exoplanet
atmosphere spectrum, mostly because there are no spectral features
that are not heavily contaminated by water vapor spectral features.
In the UV, H$_2$S absorption features are mixed in with those of
SO$_2$ and elemental sulfur, both of which are likely to be present in
the atmosphere with H$_2$S (http://www.atmosphere.mpg.de/enid/2295/)
and for specific UV cross sections see the references in
\citet{hu2012}. At visible wavelengths there are no prominent spectral
features. In the IR, at 30-80~$\mu$m, a huge amount of H$_2$S would be
needed to differentiate from water vapor spectral features.  Even so,
the H$_2$S features may cause a quasi-continuum absorption lowering
the thermal emission flux below a reasonable detection
threshhold. There might be a way to detect H$_2$S in a spectrum that
spans UV to IR: it might be plausible to detect total sulfur via UV
spectral characteristics, infer from this a relative absence of
SO$_2$, and then infer from an IR signature that there was a
significant flux of H$_2$S (or DMS) from the ground. This is
implausibly demanding of instrumentation, but not in theory
impossible.

H$_2$S is commonly regarded as a poor biosignature gas
because it is released by volcanoes.  In general, we
consider an atmospheric gas a potential biosignature gas if it is
present in such large quantities, that, in the context of other
atmospheric gases, has no likely geological origin. 

We nonetheless now turn to explore H$_2$S as a biosignature gas based
on its biomass estimate.  On a highly reduced planet, organisms could
gain energy from reducing sulfate to H$_2$S, but in this environment
H$_2$S would likely be the dominant volcanic sulfur gas as well. On an
exoplanet with a more oxidized crust and atmosphere, we can imagine an
ecosystem of sulfur disproportionators as the primary producers.

Microorganisms can disproportionate sulfur compounds of intermediate
oxidation state, including thiosulfate, sulfite, and elemental sulfur,
into H$_2$S and sulfate \citep{fins2008}.  H$_2$S would be released as
a biosignature gas.  For example, the sulfite reducers include
microorganisms in the genus Desulfovibrio and Desulfocapsa that obtain
energy from the disproportionation of sulfite \citep{kram1989} The
equation of the disproportion of sulfite in the ocean is
\begin{equation}
4 {\rm HSO_3^{-} \rightarrow 3SO_4^{2-} + 2H^+ +
H_2S}. 
\end{equation}
Note that the accumulation of sulfite and sulfate in an ocean
would make the ocean acidic, and at acidic pH levels H$_2$S will exist
in solution primarily as H$_2$S molecules, which will exchange readily
with the atmosphere (unlike a neutral or basic ocean, where S(II)
would exist as HS$^{-}$ or S$^{2-}$ which are not volatile).

To continue to explore H$_2$S as a biosignature gas, in our biomass
estimate framework, we take an Earth-size, Earth-mass planet with a
1-bar exoplanet atmosphere composed almost entirely of N$_2$ with a
minor H$_2$O concentration (like Earth above the cold trap but much
drier below the cold trap) and a small amount of CO$_2$ (1 ppm),
assuming volcanic emission.  We consider detection at IR wavelengths,
where absorption of H$_2$S might only be detected when the planet has
a relatively dry troposphere (10$^{-6}$ mixing ratio throughout the
troposphere), meaning that there will be reduced contamination of
H$_2$S spectral features by water vapor. Even for an extremely high
H$_2$S emission (3000 times Earth's volcanic emission) on a desiccated
planet, we have estimated an H$_2$S concentration of 10~ppm for
detection via thermal emission in the 45-55~$\mu$m range according to
our detection metric (\S\ref{sec-detectionmetric}). The H$_2$S surface
flux required is 3000 times Earth's volcanic emission,
($10^{17}$~molecules~m$^{-2}$~s$^{-1}$). The SO$_2$ surface flux (used
by the sulfur disproportionators) is scaled up accordingly ($2 \times$
the H$_2$S surface flux). To compute the $\Delta G$ value we used the
reactant and product molecule concentrations at the surface in terms
of partial pressures as follows SO$_2 = 2.9 \times 10^{-7}$, H$_2$S$ =
9.4 \times 10^{-6}$, and we assumed the ocean was saturated with
sulfate (1.1 mole~litre$^{-1}$) with an ocean pH of 3 (i.e., 10$^{-3}$
mole). This results in a $\Delta G = 71.3$~kJ~mole$^{-1}$.

To summarize, using equation~(\ref{eq:bio3}) and the values $P_{m_e} =
7.04 \times 10^{-6}$~kJ~g$^{-1}$~s$^{-1}$ at 290~K; $F_{\rm source} =
3.0 \times 10^{17}$~molecule~m$^{-2}$~s$^{-1}$ (or $5.0 \times
10^{-7}$~mole~m$^{-2}$~s$^{-1}$); and $\Delta G =
-71.3$~kJ~mol$^{-1}$, we find a surface biomass density of about
5.1~g~m$^{-2}$.  The total flux produced on an Earth-sized exoplanet
would be $4.0 \times 10^5$~Tg~yr$^{-1}$.  This total flux is very
high, on the order of $10^{5}$ times more than H$_2$S produced on
Earth, and similar to the benchmark values for primary production of
carbon on Earth.

A further challenge with H$_2$S as a biosignature (or even as a
geosignature) has to do with atmospheric photochemistry.  As soon as
H$_2$S (or SO$_2$) is released into the atmosphere at an amount
greater than 10 to 100 times Earth's current H$_2$S or SO$_2$ surface
flux, a blanket of aerosols or condensates form. These aerosols or
condensates are present at optically thick amounts, potentially
masking any H$_2$S or SO$_2$ spectral features, depending on the
particle size distribution.

As a side note we explain why we consider sulfur dioxide, SO$_2$, is a
failed potential biosignature gas, even though it is a gas produced by
life. SO$_2$ as a Type I biosignature would result from the
oxidation of sulfur or sulfides, via the reaction
\begin{eqnarray}
{\rm S_2^- + 3[O] + 2H^+ \rightarrow  SO_2 + H_2O} \\
{\rm S + 2[O] \rightarrow SO_2}
\end{eqnarray}
(where [O] represents an oxidizing agent such as Fe$^{3+}$ or O$_2$.)
The resulting SO$_2$ would almost certainly be dissolved in water as sulfite; in an oxidized environment further energy would be generated from oxidizing sulfite to sulfate
\begin{equation}
{\rm SO_3^{2-} + [O] \rightarrow SO_4^{2-}}.
\end{equation}
So if the environment is sufficiently oxidizing to allow energy
generation from sulfide oxidation, the end product is likely to be
sulfate, not SO$_2$. Furthermore, because SO$_2$ is geologically
generated, it would be hard to distinguish from biogenic SO$_2$.  If
there are oxidants available for life (i.e., there are oxidants
available for life to use to oxidize sulfides for energy), then the
crust and upper mantle will also be oxidized, and SO$_2$ will be a
major volcanic gas.

In summary, H$_2$S is a potential biosignature gas as seen from the
view of a biomass estimate, in a reduced atmosphere where H$_2$S can
accumulate. In general H$_2$S is ruled out as detectable or
identifiable as a biosignature gas because of its weak or
H$_2$O-contaminated spectral features and because of geological false
positives.

\subsection{CH$_3$Cl on an Earth or Early Earth Type Atmosphere}

We now turn to the Type III biosignature gases, using methyl chloride
CH$_3$Cl as the example.  On Earth CH$_3$Cl is thought to generated
mostly by land soils (by plants) \citep{kepp2005}, changed from an
earlier view that phytoplankton in the open ocean contributed most of
the Earth's CH$_3$Cl.  On Earth, CH$_3$Cl has a global production rate
of between 2-12 Tg~yr$^{-1}$ (see Table~\ref{tab-FieldFlux} and
references therein).  Averaged over Earth's land mass, the CH$_3$Cl
translates into a source flux of $3.2 \times
10^{-12}$~mole~m$^{-2}$~s$^{-1}$. This value of source flux does not
produce a detectable biosignature gas in the ``Earth as an exoplanet
spectrum'', where CH$_3$Cl has a mixing ratio of 1~ppb.  For an
Earth-like atmosphere around a low-UV quiet M dwarf, CH$_3$Cl can
accumulate up to 1~ppm \citep{segu2005}.

For an Earth-like exoplanet atmosphere spectrum, CH$_3$Cl is a
difficult biosignature gas to detect because of its overlap with other
spectral features, notably O$_3$ or CO$_2$.  As reported in
\citet{segu2005}, the CH$_3$Cl spectral feature at 9.3-10.3 $\mu$m
coincides with the O$_3$ 9.6 $\mu$m band, and would be difficult to
identify, necessitating detection of CH$_3$Cl features at 13-15 $\mu$m
or a weaker feature near 7 $\mu$m.  Under the conditions of a UV-quiet
M star with low OH concentration, \citet{segu2005} have shown that
CH$_3$Cl in an Earth-like atmosphere orbiting a quiet M dwarf can
accumulate to 1~ppm.  In an N$_2$ atmosphere with little CO$_2$,
CH$_3$Cl would be easier to detect, but still difficult owing to even
a tiny amount of atmospheric CO$_2$.

With our framework for biomass estimates, we ask the question, ``What
biomass surface density of CH$_3$Cl-producing life is required to
generate a detectable CH$_3$Cl biosignature gas?'' We answer the
question for present-day Earth and for early-Earth conditions before
the rise of atmospheric O$_2$. We take a planet of Earth's size and
mass with an Earth-mass atmosphere.  The temperature profiles are
self-consistently computed with photochemistry, and the semi-major
axis of the planet is adjusted so that the surface temperature is kept
at about 300~K.

\begin{figure}[ht]
\begin{center}
\includegraphics[scale=.33]{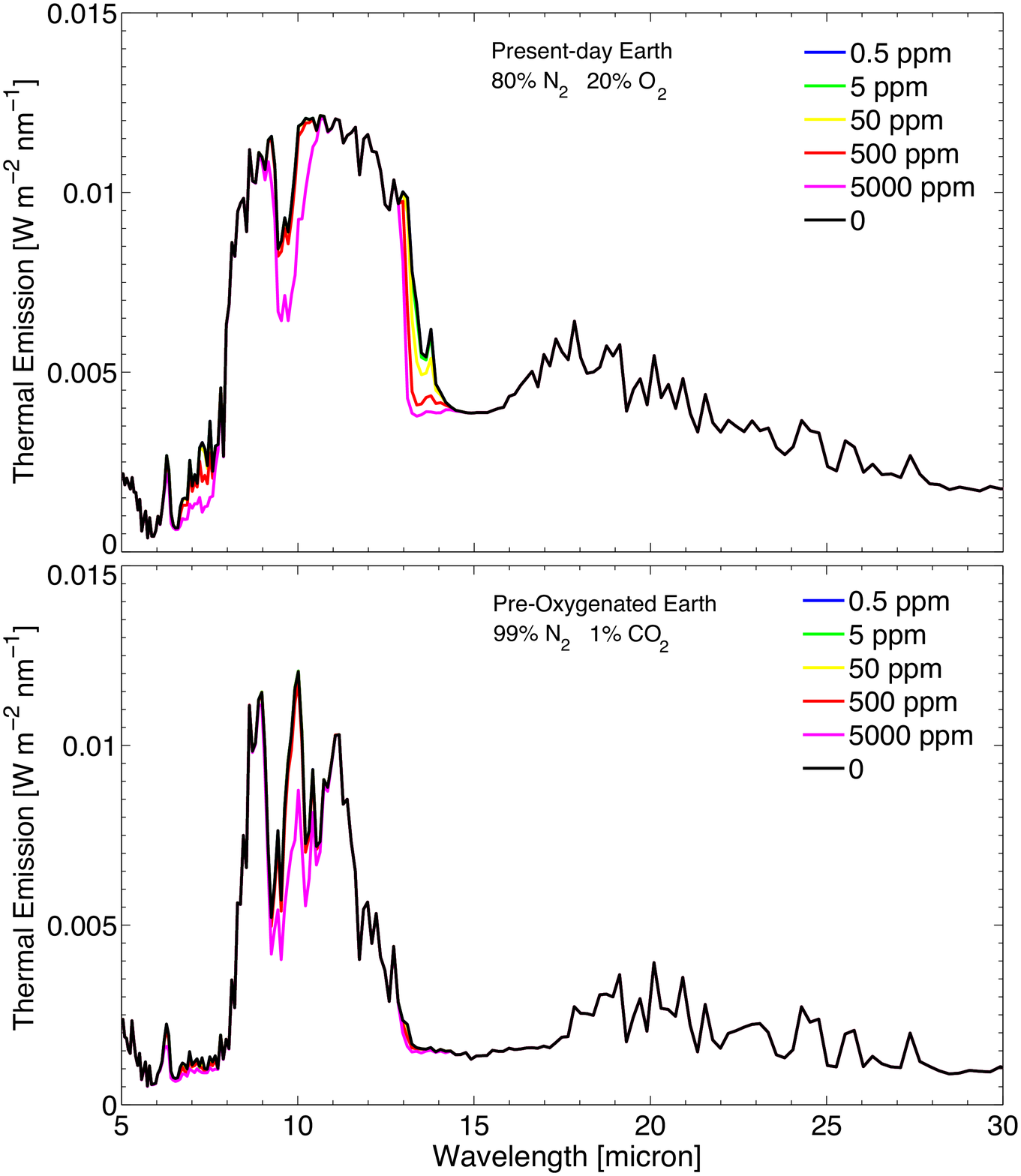}
\caption{Synthetic spectra of terrestrial exoplanets with various levels of
CH$_3$Cl. Top panel: present-day Earth's atmosphere with N$_2$, O$_2$,
H$_2$O, CO$_2$ and other photochemical derivatives. Bottom panel:
pre-oxygenated Earth's atmosphere with N$_2$, CO$_2$, H$_2$O, CH$_4$
and other photochemical derivatives.  The spectra are computed at high spectral
  resolution and binned to a spectral resolution of 100. The main 
point of this figure is how challenging it would be to detect CH$_3$Cl.
}
\label{fig:CH3Clspectra}
\end{center}
\end{figure}

For a planet with the present-day Earth atmosphere concentration, our
detection metric (\S\ref{sec-detectionmetric}) finds a 20~ppm mixing ratio at
13-14~$\mu$m on the short-wavelength wing of the 15~$\mu$m CO$_2$ band
(note that in Earth's atmosphere, the CO$_2$ concentration is low enough not to
fully saturate the 15~$\mu$m wing). The concentration of 20~ppm is 20,000
times more than Earth's current atmospheric concentration of CH$_3$Cl, a
value of 0.001~ppm. See Figure~\ref{fig:CH3Clspectra}.

The Type III biomass equation scales linearly
(equation~(\ref{eq:TypeIIIBiomassModel})), so for all other
atmospheric conditions being equal to the present-day Earth, the
biomass must therefore be 20,000 times higher than the present day.
By adopting the scaling relationship, we also assume that the 20~ppm
concentration of CH$_3$Cl is not enough to feed back on the dominant
destructing molecule [OH] concentration or to significantly effect the
temperature-pressure profile.

Is it reasonable to imagine a world with a planet biomass surface
density of CH$_3$Cl-generating life 20,000 times higher than on Earth?
If the conventional view that CH$_3$Cl is produced overwhelmingly by
oceanic plankton is adopted, then densities of tens of kg of
phytoplankton per m$^{-2}$ in the oceans would be required.  If the
more recent view that land plants are a major source of CH$_3$Cl, then
a 20,000-fold increase in biomass requires all the planet's land be
covered with plants at a density not even achieved in the most
intensively farmed land in the most favorable conditions. Neither seem
plausible. We can only escape from this conundrum if we assume that a
much larger fraction of organisms on an exoplanet produce CH$_3$Cl, or
they produce it at a much higher rate than organisms on Earth. We have
no reason for making such assumptions.

We now turn back to Earth's past, prior to the oxidation of Earth's
atmosphere, with solar-like EUV conditions, where the CO$_2$
concentration was thought to be as high as 1\% to 20\% by volume
\citep{ohmo2004}.  In an atmosphere with so much CO$_2$, CH$_3$Cl is
much harder to detect than on present-day Earth because of overlap
with the 15~$\mu$m CO$_2$ feature, the CH$_3$Cl feature at 10~$\mu$m
must be considered. A CH$_3$Cl concentration of 1000~ppm would be
needed (and barely detectable according to our detection metric), for
either the 1\% or 20\% CO$_2$ atmosphere, requiring a surface
biosignature gas source flux of $3.5 \times 10^{19}$~m$^{-2}$~s$^{-1}$
(or $5.8 \times 10^{-5}$~mole~m$^{-2}$~s$^{-1}$).  Although our
photochemistry code does not yet self-consistently treat halogentated
compounds, we computed this surface source flux by considering the
loss rates: The source flux $F_{\rm source}$, is the production rate
$P$ integrated over an atmosphere column, and $P$ is related to the
loss rate $L$, (recall \S\ref{sec-photochemmodel})
\begin{eqnarray} 
F_{\rm source} = \int_z P = \int_z {[\rm CH_3Cl]}(z) L(z) \nonumber \\
= \int_z {\rm[CH_3Cl]}(z) {\rm [OH]}(z)  K_{\rm CH_3Cl,OH}(z),
\end{eqnarray}
where [OH] is the main reactive molecule that destroys CH$_3$Cl. The
loss rate scales linearly with [OH], as long as the concentration of
CH$_3$Cl does not affect [OH]. For values of reaction rate $K$, see
Table~\ref{tab-TypeIIIKAB}.

The estimate of the surface biomass density required for a surface
source flux of $3.5 \times 10^{19}$~molecules~m$^{-2}$~s$^{-1}$ can be
found using the estimate for the Type III biosignature gases
(equation~\ref{eq:TypeIIIBiomassModel}). For CH$_3$Cl we find
$\Sigma_B = 9.4 \times 10^{5}$~g~m$^{-2}$, higher than any reasonable
biomass surface density. The uncertainties in the Type III biomass
estimate, however, should be considered.

To summarize this subsection, Type III biosignature gases are not
produced in large quantities on Earth because they are specialized
chemicals, each produced by a small number of species. Furthermore,
collisional destruction by OH is rapid. Type III 
biosignature gases have shown to be detectable in low-UV
environments, which as a consequence have less OH. We have shown that
in order to reach levels detectable on present-day Earth, we can scale
up the biomass estimates based on the concentration of the atmospheric
gas required for detection, although in the case of CH$_3$Cl this
results in an implausible biomass surface density.

\section{Discussion}
\label{sec-discussion}

\subsection{On the Order of Magnitude Nature of the Biomass Estimates}
\label{sec-orderofmag}

The biomass estimates are limited to an order of magnitude for Type I
biosignatures, and about two orders of magnitude for Type III
biosignature gases.

Estimates of the Type I biomass are dependent on $P_{m_e}$ whose
constants are known to 40\% 1-$\sigma$ \citep{tijh1993}, and $P_{m_e}$
is very sensitively dependent on temperature, due to an exponential
term (equation~(\ref{eq:Pme})). Small changes in surface temperature
estimates can have very large effects on $P_{m_e}$. Of equal
importance, is that chemical reaction rates also exponentially
increase with temperature\footnote{Chemical reaction rates often
  follow exponentials, especially reactions involving stable species,
  of the form $R \sim C \exp(T/k)$, were $R$ is the reaction rate, $C$
  is a constant, $T$ is temperature and $k$ is a constant.}.  For the
same change in temperature, the rate of destruction (and hence
production) of a gas $F_{\rm source}$, and the minimal maintenance
energy rate $P_{m_e}$ have opposite, although not necessarily
balanced, effects on our estimate of biomass, via the Type I biomass
equation~(\ref{eq:TypeIBiomassModel}).  The effect of temperature
uncertainty needs to be further investigated. A more minor
contribution to the inaccuracy of Type I biosignature gases is that
the energy released in a reaction (captured $\Delta G$) varies with
pH, and reactant and product concentrations, none of which are known
for exoplanet environments.

The Type III biomass estimate is uncertain because it is not
constrained by thermodynamics. The accuracy-limiting assumption for
the Type III biomass estimate is that exoplanet biosignature gas
production rates are the same as those found in the Earth's lab-based
maximum production rates.  

Unlike applied physics we do not know everything about biology; we do
not know everything about Earth; and we do not know everything about
atmospheric chemistry. In other words, the model is not 100\% accurate
and we must live with uncertainty in the biomass estimates. The point
is that the biomass estimate should be used to answer the question,
``Is the proposed biosignature gas plausible?'' and not for any kind
of precise prediction of biomass surface densities.

\subsection{On the Possible Terracentricity of the Biomass Estimates}
\label{sec-terracentricity}

A question arises as to the terracentricity of the biomass estimates,
and whether or not we have interchanged the conventionally used
Earth-based surface biofluxes with an Earth-based biomass model
estimate.

For the Type I biosignature gases we argue no, because thermodynamics
is universal. The Type I biosignature biomass model uses a prefactor
$A$ and an activation energy $E_A$ for the minimum maintenance energy
rate $P_{m_e}$ equation.  These parameters are lab-measured values for
Earth-based microbes. A critical question is to what extent are $A$
and $E_A$ specific to Earth life. A simple argument is that the
particular repair mechanisms and molecular turnover involved in
maintaining an organism are specific to Earth life, which is very
unlikely to be exactly replicated on other worlds.  There are stronger
arguments, however, that the energy rate is more broadly applicable:
$P_{m_e}$ follows Arrhenius' law, we think because the rate of
molecular component damage (and hence repair rate) is no different
than other chemical reactions---well described by an Arrhenius
equation.  $E_A$, the activation energy of a reaction, represents the
energy needed to break chemical bonds during the chemical reaction. In
uncatalyzed reactions, $E_A$ comes from the thermal energy of the two
reacting molecules, which itself follows from the Boltzmann velocity
distribution. The probability that any two colliding molecules will
have a combined energy greater than $E_A$, is $\exp[-E_A/RT]$. The
parameter $A$ is an efficiency factor that takes into account that
molecules have to be correctly oriented in order to react. Thus the
basic physics of the Arrhenius equation is general to all chemistry.

The argument for applying the Arrhenius equation to calculating $P_{m_e}$
starts with the point that terrestrial
life is composed of many of the possible structures in CHON
chemistry. The principle way that random chemical attack breaks those
molecules down is through hydrolysis (attack by water) or oxidation
(attack by oxygen) if oxygen is present.  The $A$ and $E_A$ terms in
the equation for $P_{m_e}$ represent those relevant to the breakage of
the most fragile of metabolites (as more stable molecules will not
need to be repaired).  Although the specific chemicals in
non-terrestrial metabolism could be quite different from those on
Earth, the nature of the chemical bonds will be similar. In
particular, non-terrestrial biochemistry will be made up of chemicals
that are moderately stable at ambient temperature and pressure, but
not too stable.  Thus the overall distribution of molecular stability
in a non-terrestrial metabolism is likely to be similar to that in
terrestrial metabolism, even if the chemical specifics differ. As a
consequence, it is reasonable to propose that the rate of breakdown of
those metabolites, and the rate at which energy is needed to repair
them (i.e., $P_{m_e}$) will be of a similar order of magnitude as the
rate of breakdown and energy requirement seen on Earth. Therefore $A$
and $E_A$ are likely to be based on chemical principles and therefore
similar to those calculated for Earth.

Turning to the terracentricity of the production rates for Type III
biosignature gases, they are derived from lab measurements of each
organism, and are likely to be specific to those organisms. The rates
may, however, be plausible indications of the Type III fluxes to be
expected from non-terrestrial life because the gas production
represents investment of energy and mass for a speciaized biochemical
function. The maximum flux rates used here represent the maximum
investment that organisms make in these Type III gases, given
essentialy unlimited energy and nutrient resources in a lab
envirionment. We speculate that it is unlikely that non-terrestrial
life would be more wasteful of resources through making any Type III
biosignature gas at rates orders of magnitude greater than those used
in this study.

The biomass surface density limits we use as a reference point are
based on Earth data, and so are terracentric. We believe that adopting
Earth values is an acceptable approximation in our model, as what
limits Earth life in environments with abundant nutrition is the
physics of mass transfer, not the specifics of how Earth life
evolved. This, however, should be validated by future research.  

We do not argue the biosignature biomass model estimates are accurate,
rather we emphasize the goal of the biomass model estimate is the
order of magnitude nature for a first order asssessment of the
plausibility of a given biosignature gas candidate.

\subsection{Biomass Estimates in the Context of an Ecology}
\label{sec-ecologycontext}

A serious criticism against the biomass estimate is lack of an ecology
context. An ecology will contain organisms that consume gases as well
as produce them. Hence the concern is that potential biosignature
gases will be destroyed by life in the same ecosystem, rendering the
biomass estimates invalid.  Indeed the biomass estimate must be a
minimum biomass estimate because there is no guarantee that
biosignature gas flux ($F_{\rm source}$) is not being consumed by
other organisms.

The biomass estimate model is intended as a check on the plausibility
of a specific gas as a biosignature gas. It is not intended to be a
prediction of the ecology of another world.  If the biomass estimate
is low, then we have confirmed that the gas is plausible as a
biosignature, given the caveats presented in \S\ref{sec-orderofmag}
and \S\ref{sec-terracentricity} and discussed throughout this paper.
In this low biomass estimate case, even if the planetary ecology has a
mix of gas-producing and gas-consuming organisms, a net production of
the gas from a moderate biomass is quite plausible. If the biomass
estimate is too high, the gas is not a plausible biosignature gas in
any ecology.  For the intermediate case where a large but not
unreasonable biomass is needed to generate a detectable biosignature,
the decision on whether the gas is a plausible biosignature is more
complicated, and will depend on the context: geochemistry, surface
conditions, atmospheric composition and other factors.

In the future when we have spectra with candidate biosignature gas
detections, in most cases we will assign a probability and not a certainty
to a biosignature gas candidate. The biomass estimate, in the context
of an ecology, will be just one of the input factors to the
probability assessment. We give an example here to sketch out other
factors to consider, which also bear weight on the ecology.  The
example is the question, how would we interpret the detection of
500~ppm CH$_3$Cl in the atmosphere of an anoxic, Earth-like planet (well
above the detectable amount according to our detection metric
and assumed future telescope capabilities)? Our biomass
model predicts that we would need a highly implausible surface biomass
density to generate such an atmospheric concentration through
metabolism. Volcanic chemistry on Earth produces traces of methyl
chloride, but only as a tiny fraction of emitted gases, making a volcanic
source seems also highly improbable. Conceivably the CH$_3$Cl could be an
industrial waste gas from a technological civilization, but in the
absence of other signs of civilization this is also improbable. In this
abstract sense, the conundrum of the intermediate biomass estiamte
has no solution, but the plausibility of a biosignature gas is
still addressable
through Baysian statistics in principle, if the prior probabilities of
the different assumptions about geochemical sources, biological
sources or technological sources can be estimated. Our biomass model
provides a numerical approach to quantifying the assumptions made
concerning the potential biological production of a biosignature
gas. Further work will integrate this into a model of our confidence
that the detection of the gas represents a detection of life.

\subsection{Massive Atmospheres and the Biomass Estimates}

In an atmosphere more massive than Earth's 1 bar atmosphere the
biomass surface density estimates could be different, depending on the
biosignature gas loss rate mechanism.  In an atmosphere where the
photochemical loss rates dominate, the biosignature source flux and
hence biomass is the same as for a less massive atmosphere.  In an
atmosphere where the deposition rate is the dominant loss mechanism,
the biosignature source flux and hence biomass surface density will
scale linearly with planetary atmosphere mass. These conclusions are
under the caveat that the surface pressure and temperature do not
cause unusual chemistry (e.g., supercritical fluid or high chemical
kinetic rates).

The biosignature source flux (i.e., production rate) is balanced by
the loss rate, as described in
equation~(\ref{eq:photochem1}). Photochemical removal is only
effective at and above the mbar pressure level in the atmosphere,
because the UV radiation typically penetrates only down to the mbar
level. The loss rate is therefore unaffected by how much atmosphere
there is below the mbar pressure level: the mbar level can be
``sitting on top of'' a small atmosphere or a very large one. Hence
the loss rate is the same (all other factors being equal) regardless
of the total mass of the atmosphere. The loss rate is balanced by the
source flux and hence the source flux needed to maintain a given
concentration of gas in an atmosphere is unaffected by the mass of the
atmosphere (assuming that there is no new, non-photochemical loss of
gas at much higher pressures). Thus, against loss from photochemistry,
the surface biomass density required to maintain a given gas
concentration is the same regardless of atmospheric mass for any
planet of a given surface area.

Loss by deposition, in contrast to loss by photochemical removal,
occurs at the planetary surface. The loss rate at the surface is
proportional to the number density of the biosignature gas at the
surface (see equation~(\ref{eq:vdep})) The number density of any
well-mixed species scales as the surface pressure for an ideal gas
atmosphere in hydrostatic equilibrium.  We can estimate the surface
gas concentration $n_{\rm ref}$ by considering a uniformly mixed
atmosphere and integrating over a vertical atmosphere column under
hydrostatic equilibrium,
\begin{equation}
\label{eq:hse1}
\int_{p_0}^{p_{\rm ref}} dp = -\int_{0}^{z_{\rm ref}} g \rho dz,
\end{equation}
where $p$ is pressure and $g$ is surface
gravity. Here we integrate from $p=0$, $z=0$ at the top
of the atmosphere down to a reference pressure $p_{\rm ref}$ at a reference altitude. Integrating the above equation, we have
\begin{equation}
\label{eq:hse2}
p_{\rm ref} = g m_{\rm atm},
\end{equation}
where $m_{\rm atm}$ is the atmospheric mass in a vertical column of
1~m$^2$ cross-sectional area.  We can therefore define $n_{\rm atm}$
by rewriting the column-integrated mass of the atmosphere in terms of
number density and the mean molecular mass of the atmosphere
$\mu_{\rm atm} m_{H}$ (where $\mu$ is the mean molecular weight and
$m_{\rm H}$ is the mass of the hydrogen atom, 
\begin{equation}
\label{eq:nestimate}
X_i n_{\rm ref} = X_i \frac{p_{\rm ref}}{kT} = \frac{g m_{\rm atm}}{kT},
\end{equation}
where $X_i$ is the mixing ratio of the gas under consideration.

We can now explain why, when surface deposition is the dominant gas
loss mechanism, a larger biosignature surface density is required as
an atmosphere mass scales up, even for a fixed atmospheric gas
concentration.  We see from equation~(\ref{eq:nestimate}) that the gas
number density at the surface scales with the surface pressure $p_{\rm
  ref}$. The deposition velocity scales with the number density and
hence surface pressure.  With a higher loss rate that scales linearly
with surface pressure, a higher source flux (production rate) is
required to balance the loss rate. Because biomass scales linearly
with source flux, a higher biomass surface density is required.

\subsection{False Positives}
\label{sec-falsepositives}
Type I biosignature gases are fraught with geologically-produced false
positives because geological processess have the same chemicals to
work with as life does. The redox reaction chemical energy gradients
exploited by life are thermodynamically favorable but kinetically
inhibited. Enzymes are used by life to accelerate the reaction. We can
be sure that a chemical reaction that is kinetically inhibited in one
environment on Earth could proceed spontaneously somewhere else on the
planet (in an environment with a favorable temperature, pressure, and
reactant concentrations). Hence Type I biosignature gases will almost
always have a possible geological origin.

Typically astronomers assume that the biosignature gas must be
produced in high enough quantities that it couldn't be confused with a
geophysical false positive. But how high of a surface flux could be
produced geologically? We plan to model the maximal geofluxes possible
for planets of different characteristics (Stamenkovic and Seager, in
prep.), with the end goal of assigning a probability to the dominant
Type I biosignature gases (H$_2$S, CH$_4$, CO, CO$_2$, N$_2$O, NO,
NO$_2$) for being produced as biofluxes vs. geofluxes. More progress
might also be made with biogeochemical cycles and the whole atmosphere
context via other atmospheric diagnostics.

False positives for O$_2$, the most obvious Type II biosignature gas
in an oxidizing environment, are limited and can most likely be
identified by other atmospheric diagnostics. For example,
photodissociation of water vapor in a runaway greenhouse with H
escaping to space could lead up to detectable O$_2$ levels. This
situation could be identified by an atmosphere heavily saturated with
water vapor. O$_2$ could also accumulate in a dry, CO$_2$-rich planet
with weak geochemical sinks for O$_2$, a case which could be
identified via weak H$_2$O features \citep{sels2002, segu2007}.

Type III biosignature gases, in contrast to Type I biosignature gases,
are less likely to have false positives. Type III biosignature gases
are chemicals produced by life for reasons other than energy capture
and are not usually naturally existing in the environment. As
byproduct gases of highly specialized physiological processes, the
Type III biosignature gases tend to be larger or more complex
molecules than Type I biosignature gases, and are not usually
replicated by non-biological processes.  In general, the more
complicated a molecule is (i.e., the more atoms it has) and the
further from fully oxidized or reduced the molecule is, the less are
produced by geological sources as compared to more simple
molecules. For example, volcanoes produce large quantities of CO$_2$,
somewhat smaller amounts of CH$_4$, small amounts of OCS, trace
amounts of CH$_3$SH, and none of isoprene.  The downside to Type III
biosignatures is that because they are usually such specialized
compounds they typically are produced in small quantities that do not
accumulate to levels detectable by remote sensing.

For Type III biosignature gases we should therefore depart from
requiring huge concentrations in the atmosphere. But, as we have seen,
detectable atmospheric concentrations are almost by definition high
concentration.

Light isotopes are used to identify biologically-produced molecules on
Earth. For exoplanets, no planned telescope will allow molecular
isotopes to be observationally distinguished from one another. In the
distant future when isotopic ratios of molecules are observable, care
has to be taken to understand the isotopic distribution of all key
molecules in the environment. In particular the isotopic ratios of the
input gases must be different than the biologically output gases. The
isotopic ratios of both the input and output gases need to be
inventoried, because exoplanets may have varying natural
distributions of isotopic ratios not seen on Earth.

\subsection{Aerosols and Hazes}

The view in biosignature gas studies is to find a biosignature gas
that exists in concentrations of orders of magnitude above the
naturally occuring values. This picture may supercede the goal to find
biosignature gases that are out of redox equilibrium (such as O$_2$
and CH$_4$) because while both might exist they might not both be in high
enough quantities to be detectable remotely.

For a biosignature gas produced orders of magnitude higher than
natural values, how much is too much? We have seen from our H$_2$S
study (\S\ref{sec-H2S}) that when H$_2$S is emitted from the surface
at an amount greater than 10 to 100 times Earth's current H$_2$S or
SO$_2$ surface flux, a blanket of aerosols or condensates form
\citep{hu2013}. These aerosols or condensates are present at optically
thick amounts, masking any H$_2$S or SO$_2$ spectral features. The
particle size partially dictates the optical properties of the
aerosols, and hence which wavelengths the spectral features will be
washed out.

Aerosol formation by CH$_4$ photolysis in N$_2$-CO$_2$ atmospheres are
expected to be significant when methane reaches 5,000 ppmv (Haqq-Misra
et al. 2008). There are two effects if aerosol formation is
significant. First, aerosols may be a net loss for CH$_4$, leading to
a decreasing marginal gain in atmospheric CH$_4$ concentration for
increasing CH$_4$ emission. Second, aerosols may impede CH$_4$
detection. As an aside we point out a basic assumption in the haze
formation model used by \citet{haqq2008} that could be improved upon:
a treatment of all hydrocarbons higher than C$_4$ as solid particles
when they might be in the gas phase makes hazes easier to form.

NH$_3$ emission itself does not lead to aerosol formation, as NH$_3$
is readily converted to N$_2$ by photolysis.

The situation for DMS is virtually the same as H$_2$S and
SO$_2$. Sulfur in the terrestrial atmosphere is likely to end up as
aerosols (S$_8$ and H$_2$SO$_4$, for H$_2$S and SO$_2$
respectively). The reason is that S$_8$ and H$_2$SO$_4$ are relatively
easy to form from sulfur gas compounds, and they have relatively low
vapor pressures enabling aerosol formation at Earth atmospheric
temperatures \citep[see][]{sein2006}.

\subsection{Subsurface Life}

If the surface of a planet is not habitable, could the subsurface
harbor life? On Earth, there is substantial subsurface life, but its
effect on the atmosphere is limited.  On Earth, surface life will
likely use any product of subsurface life as a food source, generating
the characteristic biosignature gas of the surface life. Thus H$_2$S
or CH$_4$ emitted by subsurface life on Earth is (largely) oxidized at
the surface, and so does not accumulate in the atmosphere.

If the surface is not habitable, however, then any subsurface
biological activity will eventually affect the surface, just as
surface biology on Earth eventually oxidized the stratopshere. In the
absence of surface life, subsurface life biosignature gases will diffuse to
the surface and then escape. (The methane on Mars may be an example of
this). If Mars had surface life, then that surface life would ``eat''
all the methane, and none would accumulate in the atmosphere. Thus our
model is applicable to the biomass of subsurface life as well as
surface life, with the caveat that we do not know what a plausible
upper boundary on the density of subsurface life is. 

There remains the problem that subsurface life will generate gases
that are just absorbed by the surrounding material (rock, ice, or
water). If life is too deep (as life in the internal oceans of the
Gallilean moons would be, if there is any) then biosignature gases
would take geological time to reach the surface, and would be
chemically transformed by the interposed geology (rocky or icy) in the
process.  In other words, if life is deep, the rock will only saturate
on geological timescales, and over that time any biosignature will be
chemically converted to other substances.  If life is subsurface but
shallow, rocks will become saturated with biosignature gases in a
short timescale and the biosignature may then be outgassed
to the atmosphere.

\subsection{Life on Titan: Ruled Out by Biomass Calculations?}
It has been speculated that anomalies in the atmosphere of Titan could
be signs of surface life. With substantial caveats, we can apply the
biomass model to test the plausibility of Titanian life.  Acetylene is
not detected on the surface of Titan, as reported by \citet{clar2010},
although some models suggest that acetylene should be more abundant on
the surface than benzene, which was detected.  There is an apparent
deficit of acetylene on Titan's surface, because acetylene was
detected in Titan's atmosphere. \citet{stro2010} modeled the
atmosphere of Titan, and predicted a tropospheric deficit of H$_2$ in
Titan's atmosphere, compared to stratospheric levels, implying a
downward flux of H$_2$ of the order of $2 \times
10^{14}$~m$^{-2}$~s$^{-1}$.  Several authors have speculated this is a
sign of life on Titan \citep[see
  e.g.,][]{norm2011,seag2012}\footnote{see also
  http://www.ciclops.org/news/making\_sense.php?id=6431\&js=1},
deriving energy from either of the following reactions,
\begin{eqnarray}
{\rm C_2H_2 + 2H_2 \rightarrow C_2H_6} \hspace{0.1in}
\Delta G_0 = -242 \hspace{0.02in} {\rm kJ~mole^{-1}} \\
{\rm C_2H_2 + 3H_2 \rightarrow 2CH_4 \hspace{0.1in}} 
\Delta G_0 = -375 \hspace{0.02in} {\rm kJ~mole^{-1}}. 
\end{eqnarray}

We can test the hypothesis that the acetylene deficit is a biosignature
gas using the Type I biomass model. Life on Titan could live on the
surface, using liquid methane as a solvent \citep{bain2004, mcka2005},
or near the surface, or in the deep interior using water as a
solvent. Life in liquid methane/ethane at $\sim$100~K must have
radically different chemistry from terrestrial life; it is almost
inevitable that the biomass model in
equation~(\ref{eq:TypeIBiomassModel}) will not be valid for such
different biochemistry. Specifically, we might expect the constant $A$
term in the minimal maintenance energy rate equation~(\ref{eq:Pme}) to
be lower for liquid methane life. The major source of damage for
terrestrial biomolecules is attack by water, which
will be much slower when the molecules are not dissolved in water. 
While we have no idea how
much smaller $A$ should be, we can, however, still attempt to apply
equation~(\ref{eq:bio3}) to predict the minimum biomass necessary to
generate a biosignature gas. If on Titan the $P_{m_e}$ constant $A$ is
smaller than on Earth, then from equation~(\ref{eq:Pme}), the minimal
maintenance energy rate $P_{m_e}$ will be smaller (at a given
temperature), and so from equation~(\ref{eq:TypeIBiomassModel}) our
biomass estimate will be larger.

If hydrogen and acetylene are being consumed by water-based life on
Titan today, then that life must be near the surface. Water near the
surface would freeze to a eutectic of whatever solutes are present in
the internal ocean---as these are unknown, we have assumed a
water/ammonia eutectic with a freezing point of 176~K \citep{leli2002}.
The deeply buried ``internal ocean'' is
likely to be warmer, but is too deeply buried to account for a high
surface flux of hydrogen \citep{fort2012};
however we include a calculation for a saturated freezing brine at
252~K for comparison.

The flux of H$_2$ downwards is proposed to be $2 \times
10^{14}$~m$^{-2}$~s$^{-1} = 3.3 \times
10^{-10}$~mole~m$^{-2}$~s$^{-1}$.  The surface concentration of
acetylene is taken to be 0.15~mM in water \citep{mcau1966}, in the
methane/ethane lakes acetylene concentration is taken to be the same
as its mixing ratio in the higher atmosphere \citep{stro2010}. Hydrogen,
methane, and ethane are assumed to be in equilibrium with the
atmosphere. Given these constraints, the biomass calculations for the
three conditions mentioned above and the two reactions are given 
in Table~\ref{tab-Titan}.

\begin{table*}[ht]
\begin{center}
\begin{tabular}{| l | l | l | l | l |}
\hline 
Environment &$T$ (K) & $\Delta G$ & $P_{m_e}$ & Biomass \\
& & (kJ mole$^{-1}$) & (kJ g$^{-1}$ s$^{-1}$) & (g m$^{-2}$)\\
\hline
C$_2$H$_2$ + 2H$_2$ $\rightarrow $ C$_2$H$_6$ &  & & & \\ 
\hline
Liquid methane/ethane & 100 & 298 & 1.4 $\times 10^{-30}$ & 7.1$\times 10^{22}$ \\
Ammonia/water eutectic & 176 & 288 & 6.3$\times 10^{-15}$ & 1.5$\times 10^{7}$\\
Freezing brine & 252 & 277 & 1.0$\times 10^{-8}$ & 9.0 \\
\hline 
C$_2$H$_2$ + 3H$_2$ $\rightarrow $ 2CH$_4$ &  & & &\\
\hline
Liquid methane/ethane & 100 & 385 & 1.4$\times 10^{-30}$ & 9.1$\times 10^{22}$ \\
Ammonia/water eutectic & 176 & 421 & 6.3$\times 10^{-15}$ & 2.1$\times 10^{7}$ \\
Freezing brine & 252 &  403 & 1.0$\times 10^{-8}$ & 14 \\
\hline 
\end{tabular}
\end{center}
\caption{Biomass surface density estimates for Titan in different surface environments.}
\label{tab-Titan}
\end{table*}

The values for the biomass predicted for life in liquid methane/ethane
are clearly far too high to be in any way acceptable or plausible.  We
therefore have reason to doubt that life that uses chemistry similar
to terrestrial life in liquid methane is generating the hydrogen
deficit on Titan. An obvious caveat is that equation~(\ref{eq:bio3})
based on terrestrial, carbon/water-based life: life operating at 100~K
will have radically different, and probably more fragile, chemistry
\citep{bain2004}, and hence different constants in
equation~(\ref{eq:Pme}).

Life in near-surface water ``only'' requires a Titan-covering layer
between $\sim$1 and 1.5~m thick, equivalent to a modern cabbage farm.
A Titan-wide layer of life 1.5~m thick implies a near-surface water
layer of at least this thickness across the whole moon, or a thicker
layer concentrated in specific regions of the moon, One would have
thought that evidence of this would have been detected by IR
spectrometry, which it is not \citep{clar2010}. Again our model
suggests that near-surface life is not the sink for atmospheric
hydrogen.

Only a modest density of living matter is needed to explain the
hydrogen flux if life is present in freezing brine. However if freezing
brine is present, it will be buried under a 100~km thick layer of
ice. It is unlikely that gases could exchange with the atmosphere
through an ice shell of this sort fast enough to explain the apparent
deficit of hydrogen in Titan's atmosphere.

This speculative application of the biomass model illustrates that the
model can be used to rule out Earth-like life in some circumstances
that are quite unlike Earth. As noted in \S\ref{sec-terracentricity},
there is good reason for our model to apply to other biochemistries
based on C, O, N, P, and S. If life on Titan is based on radically
different chemistry that Earth's biochemistry, then the constants
in equation~(\ref{eq:Pme}) will be different, and our model will not
accurately predict biomass requirements.

\section{Summary}
\label{sec-summary}

We have created a framework for linking biosignature gas detectability
to biomass surface density estimates.  This enables us to consider different
environments and different biosignature gases than are present on
Earth. This liberates predictive atmosphere models from requiring
fixed, terracentric biosignature gas source fluxes. We have validated
the models on terrestrial production of N$_2$O/NO, H$_2$S, CH$_4$, and
CH$_3$Cl. We have applied the models to the plausibility of NH$_3$ as
a biosignature gas in a reduced atmosphere, to CH$_4$ on early Earth
and present day Mars, discussed H$_2$S as an unlikely biosignature
gas, and ruled out CH$_3$Cl as a biosignature gas on Earth or early
Earth.

We presented a biosignature gas classification (described in
\S\ref{sec-biosigclass}), needed as a precursor to develop
class-specific biomass model estimates.  The relevant summary point is
that Type I biosignature gases---the byproduct gases produced from
metabolic reactions that capture energy from environmental redox
chemical potential energy gradients---are likely to be abundant but
always fraught with false positives. Abundant because they are created
from chemicals that are plentiful in the environment. Fraught with
false positives because not only does geology have the same molecules
to work with as life does, but in one environment where a given redox
reaction will be kinetically inhibited and thus proceed only when
activated by life's enzymes, in another environment with the right
conditions (temperature, pressure, concentration, and acidity) the
same reaction might proceed spontaneously. In contrast to Type I
biosignature gases, Type III biosignature gases---as chemicals
produced by life for reasons other than energy capture or the
construction of the basic components of life---are generally expected
to be produced in smaller quantities, but will have a wider variety
and much lower possibility of false positives as compared to Type I
biosignature gases.  These qualities are because Type III biosignature
gases are are produced for organism-specific reasons and are highly
specialized chemicals not directly tied to the local chemical
environment and thermodynamics.

Model caveats are related to the order of magnitude nature of the
biomass estimates, the possible terracentricity of the biomass model
estimates, and the lack of ecosystem context.

Exoplanets will have planetary environments and biologies
substantially different from Earth's, an argument based on the
stochastic nature of planet formation and on the observed variety of
planet masses, radii, and orbits. The biomass model estimates are
intended to be a step towards a more general framework for
biosignature gases, enabling the move beyond the dominant terracentric
gases. We hope this new approach will help ensure that out of the
handful of anticpated potentially habitable worlds suitable for
followup spectral observations, we can broaden our chances to identify
an inhabited world.

\acknowledgements{We thank Foundational Questions Institute (FQXI) for
  providing funding for the seeds of this work many years ago. We
  thank Matt Schrenk and Bjoern Benneke for useful discussions.
  W.~B. thanks Carl and Barbara Berke for support.}

\clearpage

\begin{appendix}
\setcounter{table}{0}
\renewcommand*\thetable{\Alph{section}.\arabic{table}}
\section{Terrestrial Flux References}

\begin{table}[ht]
\begin{center}
\begin{tabular}{| l | l|}
\hline
Molecule  &  Ref \\
\hline 
CH$_3$Cl  &  1-4 \\
COS       &  5 \\
CS$_2$    &  5 \\
DMS       &  6 \\
H$_2$S    &  5 \\
isoprene  &  7  \\
N$_2$O    &  8 \\
NH$_3$    &  -- \\
\hline 
\end{tabular}
\end{center}
\caption{References for field fluxes listed in Table~\ref{tab-FieldFlux}.
(1) \citet{moor1996}; (2) \citet{dimm2001};
(3) \citet{cox2004}; (4) \citet{wang2006}; (5) \citet{anej1989b}; 
(6) \citet{morr1990}; (7) \citet{fuen1996};
(8) \citet{nyka1995}. }
\label{tab-FieldFluxRefs}
\end{table}

\begin{table}[ht]
\centering
\begin{tabular}{| l | l | l | l| l| l|}
\hline
Molecule & Ref \\
\hline
N$_2$O   & 1--12 \\
NO       & 1, 4, 8, 9, 13, 14\\
H$_2$S   &  15--33\\
CH$_4$   &  34--42\\
\hline
CH$_3$Br & 43--51\\
CH$_3$Cl & 44, 46--53\\
COS      & 54--55\\
CS$_2$   & 56\\
DMS      & 55, 57--63\\
isoprene & 64--73\\
\hline 
\end{tabular}
\caption{
References for laboratory fluxes listed in Table~\ref{tab-LabRates}.
(1) \citet{kest1997}, (2)\citet{remd1991}, (3) \citet{abou1985}, 
(4) \citet{ande1986}, (5) \citet{samu1988}, (6) \citet{vorh1997}, 
(7) \citet{kasp1982}, (8) \citet{kesi2006}, (9) \citet{ande1993}, 
(10) \citet{wrag2004}, (11) \citet{shaw2006}, (12) \citet{gore1980},
(13) \citet{lips1981}, (14) \citet{schm1997}, (15) \citet{camp2001},
(16) \citet{esco2007}, (17) \citet{stet1983}, (18) \citet{para1987},
(19) \citet{bott2001}, (20) \citet{belk1985}, (21) \citet{slob2012},
(22) \citet{hube1987}, (23) \citet{brow1989}, (24) \citet{belk1986},
(25) \citet{fard1996}, (26) \citet{fins1998}, (27) \citet{jack2000},
(28) \citet{bak1987a}, (29) \citet{bak1987b}, (30) \citet{habi1998},
(31) \citet{widd1983}, (32) \citet{wall1995}, (33) \citet{boll2001},
(34) \citet{pate1977}, (35) \citet{pate1978}, (36) \citet{zeik1975},
(37) \citet{zind1984}, (38) \citet{mull1986}, (39) \citet{penn2000},
(40) \citet{pers1981}, (41) \citet{scho1985}, (42) \citet{taka2008},
(43) \citet{latu1995}, (44) \citet{dail2007}, (45) \citet{saem1998},
(46) \citet{bake2001}, (47) \citet{scar1998}, (48) \citet{latu1998},
(49) \citet{manl1987}, (50) \citet{scar1996}, (51) \citet{brow2010},
(52) \citet{tait1995}, (53) \citet{harp1985}, (54) \citet{grie1994},
(55) \citet{geng2006}, (56) \citet{xie1999}, (57) \citet{caro1994},
(58) \citet{baum1994}, (59) \citet{matr1995}, (60) \citet{anse2001},
(61) \citet{mali1998}, (62) \citet{stef1993}, (63) \citet{gonz2003},
(64) \citet{hewi1990}, (65) \citet{kess1999}, (66) \citet{broa2004},
(67) \citet{mons1994}, (68) \citet{wagn1999}, (69) \citet{shaw2003},
(70) \citet{shar1993}, (71) \citet{loga2000}, (72) \citet{fang1996},
(73) \citet{harl1996}. }
\label{tab-LabRatesRefs}
\end{table}

\end{appendix}

\clearpage
\bibliography{planets}

\end{document}